\documentclass{aa} 
\usepackage{natbib}
\bibpunct{(}{)}{;}{a}{}{,} 

\usepackage{pifont}

\usepackage{txfonts}
\usepackage[pdfstartview=FitH,      
            breaklinks=true,        
            bookmarksopen=true,     
            bookmarksnumbered=true, 
            ]{hyperref}
\usepackage{aas_macros}
\usepackage{float}
\usepackage{wasysym} 
\usepackage{subfigure}
\usepackage{placeins}
\usepackage{xcolor}
\usepackage[version=4]{mhchem}
\newcommand{\comment}[1]{}
\usepackage{ulem}
\usepackage{multirow}
\usepackage{makecell}
\usepackage{graphicx}

\begin{document}

   \title{LavAtmos 2.0}

   \subtitle{Incorporating volatile species in vaporisation models}

   \author{C. P. A. van Buchem \inst{1,2}
        \and
        M. Zilinskas\inst{3}
        \and
        Y. Miguel\inst{1,3}
        \and
        W. van Westrenen\inst{2}
          }

   \institute{
    Leiden Observatory, Leiden University, P.O. Box 9513, 2300 RA Leiden, The Netherlands \email{vanbuchem@strw.leidenuniv.nl}
    \and
    Department of Earth Sciences, Vrije Universiteit Amsterdam, De Boelelaan 1085, 1081 HV Amsterdam, The Netherlands
    \and
    SRON Netherlands Institute for Space Research, Niels Bohrweg 4, 2333 CA Leiden, The Netherlands
}


   \date{Received DD MM, 202Y; accepted DD MM, 202Y}

 
  \abstract
   {Due to strong irradiation, hot rocky exoplanets are able to sustain lava oceans. Direct interaction between these oceans and overlying atmospheres can provide insight into planetary interiors. In order to fully understand how the composition of the atmosphere of such planets are affected by the properties of the oceans, comprehensive chemical equilibrium models are required. Thus far, most models have only taken non-volatile species into account when calculating lava vaporisation.}
   {We investigate the effect of including C-, H-, N-, S-, and P-bearing species in the equilibrium lava vaporisation calculations on the overall atmospheric composition of hot rocky exoplanets by expanding our LavAtmos code.}
   {We calculated the O$_2$ partial pressure which satisfies both the laws of mass action and mass conservation in a system that contains both melt species and volatile elements. We integrated the chemical equilibrium code FastChem to expand the number of considered gas phase species to 523. We applied this new approach to calculate the composition of “pure” atmospheres which contain only C, H, N, S, or P and of more complex atmospheres which contain all five aforementioned elements. We also tested two proposed compositions for the atmosphere of 55-Cnc e.}
   {We find that the inclusion of volatile elements in vaporisation calculations increases the partial pressures of vaporised species (SiO, SiO$_2$, Na, and K) compared to volatile-free vaporisation for all tested atmospheric compositions. Moreover, including volatile species in the vaporisation reactions leads to a significantly greater O abundance in the atmosphere than in the volatile-free vaporisation case, which influences partial pressures of key volatile species such as CO$_2$ and H$_2$O. When testing the compositions proposed for 55-Cnc e, we find that a low C/O ratio could potentially serve as an indication of the presence of a surface lava ocean on an ultra-short-period planet with a volatile atmosphere.}
   {Volatile elements must be taken into account for comprehensive modelling of vaporisation from a surface lava ocean into a volatile atmosphere.} 

   \keywords{planets and satellites: terrestrial planets; planets and satellites: composition;  planets and satellites: atmospheres; planets and satellites: surface; astrochemistry}

   \maketitle

\section{Introduction}\label{sec:intro}

Recent years have seen a growing interest in the properties of ultra-short period (USP) rocky planets. The proximity of these planets to their host star causes their surfaces to be highly irradiated, reaching temperatures that exceed those required to initiate rock melting  ($\sim$1500 K). As a result a significant part of their surface could be covered by a lava ocean \citep{leger_transiting_2009,henning_highly_2018,boukare_deep_2022}. A lava ocean serves as a direct interface between the interior and the atmosphere of these planets, allowing the two reservoirs to thermochemically equilibrate with each other, creating a strong link between the compositions of the two. Hence, characterising the chemical composition of a USP rocky world's atmosphere could in principle be used to place constraints on its interior chemical composition. Thanks to the strong observational biases favouring the discovery of short period planets with current detection methods \citep{beatty_predicting_2008}, a large number of potential targets for detailed study exists \citep{zilinskas_observability_2022}. The advent of JWST and the development of telescopes such as Ariel and the ELT promise to provide the opportunity to begin characterising the chemical composition of USP atmospheres. Therefore, understanding of the interior-atmosphere compositional link of these planets is more relevant than ever and could be used to provide insights into their chemical and formation history \citep{schaefer_chemistry_2009, morbidelli_challenges_2016, lichtenberg_redox_2021}. 

To be able to infer the interior composition from the atmospheric composition of rocky planets, a quantitative understanding of outgassing and ingassing processes is required. A growing number of studies have shown that the vaporisation of a lava ocean with chemical compositions similar to silicate compositions found in our solar system should lead to atmospheres containing species such as SiO, SiO$_2$, Na, K, MgO, TiO, and a wide variety of other compounds \citep[e.g.][]{schaefer_thermodynamic_2004,miguel_compositions_2011,ito_theoretical_2015,kite_atmosphere-interior_2016,fegley_chemical_2023,van_buchem_lavatmos_2023,wolf_vaporock_2023}. The abundances of these species present in the atmosphere determine their observability, with SiO and SiO$_2$ being the easiest to detect due to their high molecular opacity between 7-10 micron \citep{zilinskas_observability_2022}. 

In much of the previous work, the simplifying assumption is made that volatile elements such as carbon (C), hydrogen (H), nitrogen (N), sulfur (S), and phosphorus (P) are absent in the atmospheres of USP rocky planets due to the extreme irradiation undergone by these planets and the resulting atmospheric erosion and escape. However, there are indications that magma oceans may potentially harbour large reservoirs of water \cite{hirschmann_magma_2012,lebrun_thermal_2013,dorn_hidden_2021,kite_water_2021}. In addition, \cite{herbort_atmospheres_2020} shows the possibility of stable hot atmospheres containing C, H, N, and S species. The emission spectrum of 55-Cnc e recently observed by JWST shows a day-side temperature of 2000 K \citep{hu_secondary_2024}, low relative to the expected equilibrium temperature. This suggests strong heat redistribution to the night side of the planet. Together with infrared absorption features at $\simeq$4.5 micron, due to either CO or CO$_2$, this suggests that this USP rocky planet supports a significant volatile atmosphere. Hence, both theory and observations are pointing to the necessity of quantitatively assessing the effects of the presence and abundance of volatile elements when modelling lava ocean outgassing on USP rocky-planets. 

\cite{piette_rocky_2023} and \cite{zilinskas_observability_2023} have shown that volatile atmospheres may suppress the signatures of vaporised species in the spectra of USP-rocky planets, but these studies do not take into account the effect that volatiles have on the vaporisation of lava and the abundances of the outgassed species. Instead, calculated abundances of the elements coming from vaporised lava are simply added to the total assumed abundances of one or more volatiles in the atmosphere. This does not take into account how volatile species influence the vaporisation reactions themselves.

\cite{charnoz_effect_2023} have investigated what the effect is of the presence of H in the atmosphere of a USP rocky-planet on the vaporisation from a lava ocean by including the formation of H$_2$O from H$_2$ and O$_2$ in chemical equilibrium calculations. They conclude that even a small amount of H ($\simeq$1 bar) in the atmosphere can increase the partial pressure of vaporised species (such as SiO and SiO$_2$) by several orders of magnitude. Using this approach, \cite{falco_hydrogenated_2024} have shown that hydrogenated atmospheres can still contain sufficient SiO ($\simeq 4$ bar at a surface temperature of 3000 K) to yield detectable SiO features in the infrared. In this paper we expand upon the approach of \cite{charnoz_effect_2023} by including all melt species available in LavAtmos \citep{van_buchem_lavatmos_2023}, as well as all the gas species included in thermochemical equilibrium code FastChem \citep{stock_fastchem_2018}. 

We first apply the resulting code (LavAtmos 2.0) to `pure' C, H, N, S, and P atmospheres in order to isolate the individual effects that these most common volatile elements have on vaporisation from a lava ocean. We then apply the code to `complex' volatile atmospheres which are dominated by one volatile element but which contain all of the other volatile elements as well. Finally, we test the effects of N$_2$, CO, and CO$_2$-rich atmospheres that are commonly assumed in studies predicting the expected spectral signature of rocky exoplanet atmospheres and that give a good fit to the recently published 55-Cnc e data \citep{hu_secondary_2024}. Based on these results, we point out which chemical species could, if detected in the atmosphere of a USP rocky planet, hint at the presence of a lava ocean and potentially constrain aspects of its chemical composition. 

\section{Methods}\label{sec:methods}

\begin{figure}
    \centering
    \includegraphics[width=\linewidth]{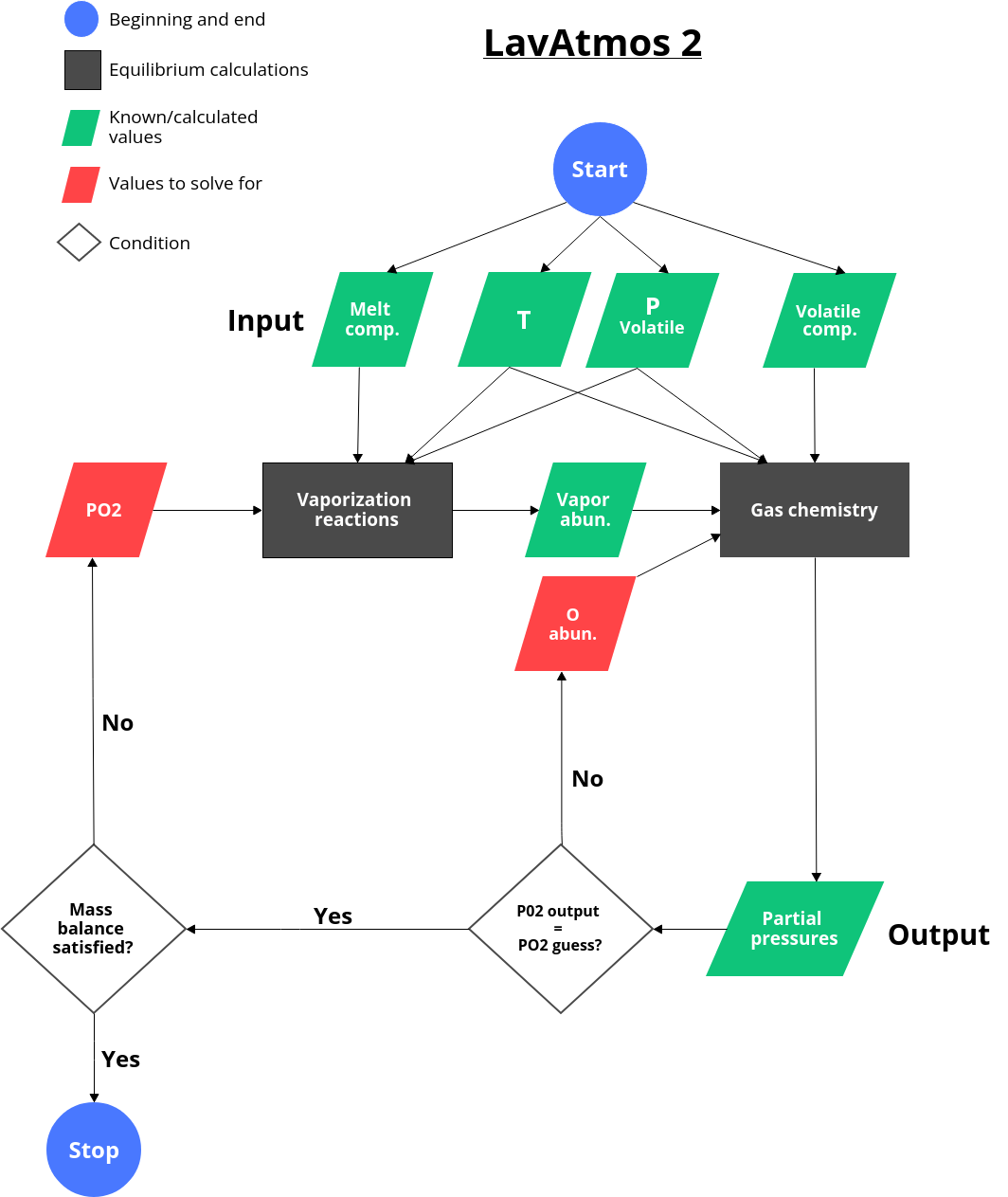}
    \caption{Diagram of the LavAtmos 2.0 equilibrium code: The given input (top of the diagram) includes the melt composition, temperature and pressure at the surface of the melt, and the elemental abundances of the volatile elements in the atmosphere. The output consists of partial pressures of all of the species in the atmosphere. The vaporisation reactions are calculated in the same way as in \cite{van_buchem_lavatmos_2023}. See section \ref{sec:methods} for a more detailed explanation of how the code works.}
    \label{fig:code_diagram}
\end{figure}

\subsection{Workflow}

LavAtmos is an open-source thermochemical vaporisation Python code described in detail in \cite{van_buchem_lavatmos_2023}. It is based on the principle of using the laws of mass action and mass conservation to constrain the partial pressure of O$_2$ (P\textsubscript{O$_2$}) released in lava vaporisation reactions. The thermodynamic properties of the lava are quantified using the MELTS code \citep{ghiorso_chemical_1995,asimow_algorithmic_1998,ghiorso_pmelts_2002,gualda_rhyolite-melts_2012,ghiorso_h2oco2_2015}. The original version of LavAtmos covers 31 gas phase species. It does not consider gas-gas reactions, and the only source of atmospheric components is evaporation from lava as calculated from liquid-gas reactions. In LavAtmos 2.0, we expanded the original concept by making use of the gas-phase chemical equilibrium solving abilities of FastChem 3 \citep{kitzmann_fastchem_2023} in order to include the effect of volatile elements (for example, those present in a primary atmosphere) on the vaporisation of melt species. We note that condensation reactions, which FastChem 3 can incorporate, are not considered in this work. 

To illustrate our method, we walk through the model assuming a lava ocean that is composed of pure molten SiO$_2$, and that only the volatile element H is present in the atmosphere. We make use of the flow-chart shown in Figure \ref{fig:code_diagram} to visualise the working of the code. Code input includes a melt composition, a surface temperature T, a total volatile pressure P, and mole fractions of the included volatile elements (green parallelograms at the top of the flow-chart). The total volatile pressure is the pressure of all volatile elements (C, H, N, S, and P) assumed to be in the atmosphere and does not include the species vaporised from the lava ocean. The total pressure of the atmosphere at the gas-melt interface is the sum of the total volatile pressure and the total vaporised pressure. The volatile composition is given as input in the form of elemental fractions of the volatile atmosphere.

With this input, the code calculates the elemental abundances of the vaporised elements coming from the melt. This is done by first calculating the partial pressures of vaporised species. Using our example of a pure SiO$_2$ melt, we have to consider the following vaporisation reactions:
\begin{equation}
    \text{SiO}_2\text{(l)} \leftrightarrow \text{SiO}_2\text{(g)},
\end{equation}
and
\begin{equation}
    \text{SiO}_2\text{(l)} - \frac{1}{2}\text{O}_2\text{(g)} \leftrightarrow \text{SiO(g)}.
\end{equation}
The partial pressures of SiO$_2$(g) and SiO(g) are calculated using the law of mass action and by assuming an ideal gas as follows: 
\begin{equation}\label{eq:PSiO2}
    \text{P}_{\text{SiO}_2\text{(g)}} = K_{1}(T)a_{\text{SiO}_2\text{(l)}},
\end{equation}
and
\begin{equation}\label{eq:PSiO}
    \text{P}_{\text{SiO(g)}} = K_{2}(T)a_{\text{SiO}_2\text{(l)}}\text{P}_{\text{O}_2}^{-1/2},
\end{equation}
where P\textsubscript{SiO$_2$(g)} and P\textsubscript{SiO(g)} are the partial pressures of the vaporised species, $K_{1}$ and $K_{2}$ are the temperature dependent chemical equilibrium constants specific to reactions (1) and (2), respectively, and $a$\textsubscript{SiO$_2$(l)} is the chemical activity of SiO$_2$ in the melt (see \cite{van_buchem_lavatmos_2023} for a more in depth derivation). The chemical equilibrium constant of each reaction is determined using the JANAF tables \citep{chase_nist-janaf_1998} and the activities of the melt species are calculated\footnote{Note that in our particular example, the activity of SiO$_2$ in the melt would be equal to 1.} using MELTS \citep{ghiorso_chemical_1995,asimow_algorithmic_1998,ghiorso_pmelts_2002,gualda_rhyolite-melts_2012,ghiorso_h2oco2_2015}. MELTS requires pressure as input due to the fact that beyond 100 bar, increasing pressure starts to have a significant effect on the activities of the endmember-species. Vaporised atmosphere pressures do not reach these pressures within the parameter space explored in this work. However, total volatile pressures do. For this reason we use the total volatile pressure as the total pressure given as input to MELTS (as can be seen in the flowchart in Figure \ref{fig:code_diagram}). 

This leaves only P\textsubscript{O$_2$} as an unknown in equation \ref{eq:PSiO}. Since we do not know a priori the correct value for P\textsubscript{O$_2$}, we make an initial guess (as indicated in the red parallelogram in the top left corner of the flow-chart in Figure \ref{fig:code_diagram}). From the resulting SiO and SiO$_2$ partial pressure values, we calculate the initial Si abundance in the atmosphere. This is passed on to FastChem 3, along with the abundance(s) of the volatile element(s), which in the case of this working example is H. The addition of H to the gas equilibrium chemistry means that the P\textsubscript{O$_2$} calculated by FastChem will not be the same as the P\textsubscript{O$_2$} used to calculate the Si abundances used by FastChem. We must therefore find the O abundance that corresponds to the P\textsubscript{O$_2$} value used for the vaporisation calculations in order to match the P\textsubscript{O$_2$} imposed by the melt with that in the atmosphere. This approach is necessary due to the fact that we cannot fix PO$_2$ a-priori when doing FastChem calculations. To achieve this, we guess an initial atmospheric O abundance (red parallelogram in the middle of the flow chart). We pass this to FastChem 3, which outputs partial pressures (green parallelogram, bottom right). If the P\textsubscript{O$_2$} output by FastChem 3 is not the same as the P\textsubscript{O$_2$} used for the vaporisation reactions, then we assign a new value for the atmospheric O abundance. This is iterated until the FastChem 3 output for P\textsubscript{O$_2$} matches the original P\textsubscript{O$_2$} guess used for the vaporisation reactions. This ensures that the law of mass action is upheld. Once this condition has been satisfied, we assess if the law of mass balance is satisfied. Assuming that the only species included in our example model are SiO, SiO$_2$, O$_2$, H$_2$, and H$_2$O, the following equation must hold:
\begin{equation}\label{eq:mb}
    0 = \text{P}_{\text{O}_2} + \frac{1}{2}\text{P}_{\text{H}_2\text{O}} - \text{P}_{\text{SiO}_2} - \frac{1}{2}\text{P}_{\text{SiO}}.
\end{equation}
This ensures mass balance for oxygen. If condition \ref{eq:mb} is not satisfied, we return to the top left of the diagram and try a new P\textsubscript{O$_2$} value. This `outer' loop as well as the `inner' loop (used to determine the correct O abundance to give to FastChem 3) are solved using the Scipy optimization function fsolve\footnote{https://docs.scipy.org/doc/scipy/reference/generated/scipy.optimize. fsolve.html.} 

In this worked example, solving for P\textsubscript{O$_2$} yields the partial pressures of the above mentioned species (SiO, SiO$_2$, O$_2$, H$_2$, and H$_2$O) for the given input parameters. In the full version of the code, partial pressures are given for all species with elements present in the melt composition and/or the volatile composition, and that are included in FastChem 3 \citep{kitzmann_fastchem_2023}. Both LavAtmos 1 and 2 are available on GitHub\footnote{https://github.com/cvbuchem/LavAtmos} under the GNU General Public License v3.

It should be noted that this method works under the important assumption that all of the O in the atmosphere comes from vaporisation reactions. This is a commonly used assumption for work done on modelling vaporisation reactions from melts into volatile-free atmospheres \citep{fegley_vaporization_1987,schaefer_thermodynamic_2004,kite_atmosphere-interior_2016,van_buchem_lavatmos_2023,wolf_vaporock_2023,charnoz_effect_2023,seidler_impact_2024}. This assumption holds if we assume that the lava ocean is the dominant oxygen reservoir with respect to the atmosphere. We also do not take the solubility in a melt of volatile species such as H$_2$O or CO$_2$ into account. The possible effects of this are discussed in section \ref{sec:discussion_water_effect}.

\begin{figure}
    \centering
    \includegraphics[width=\linewidth]{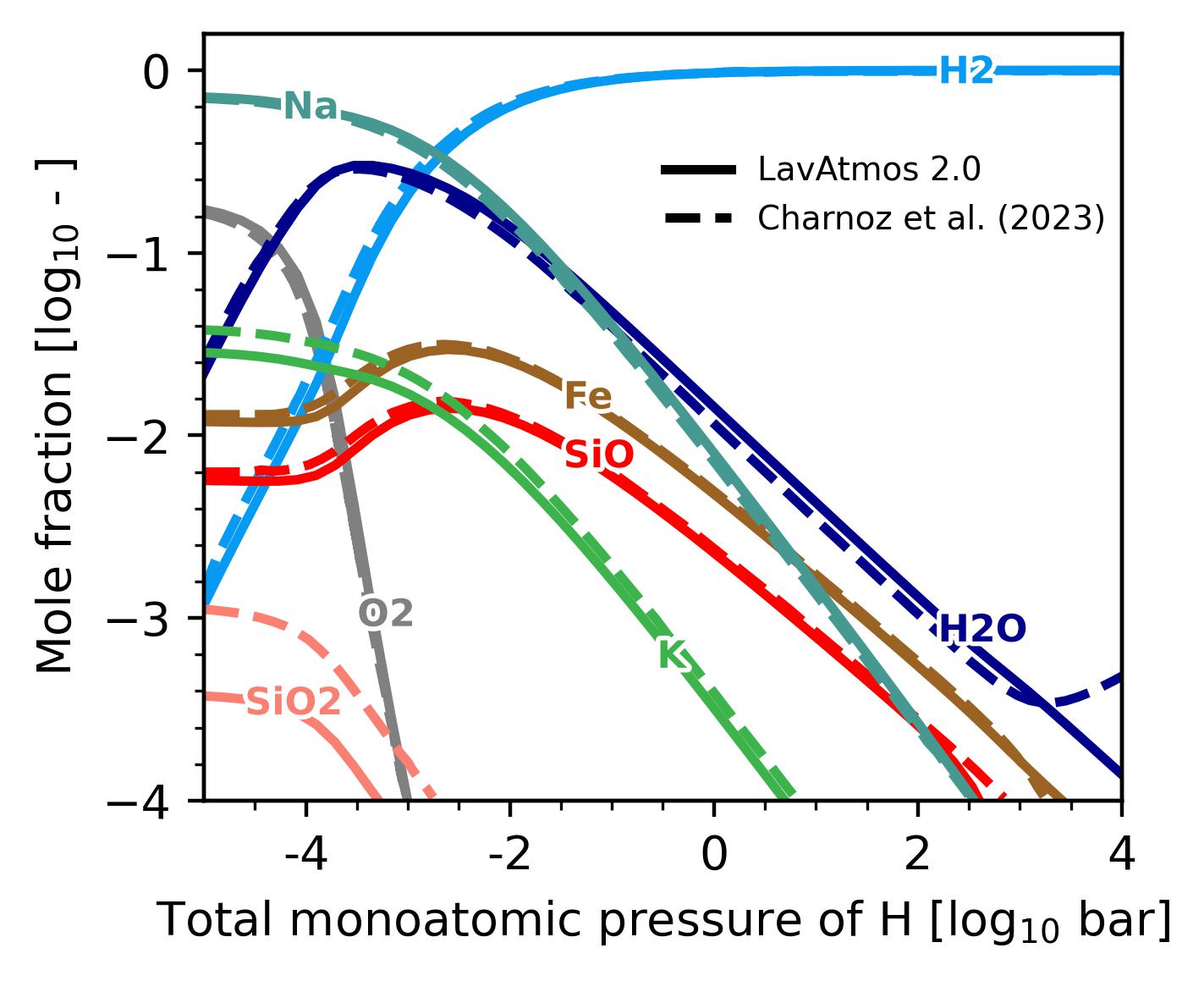}
    \caption{Mole fraction output comparison: Results from LavAtmos 2.0 are shown using solid lines and results from \cite{charnoz_effect_2023} are shown using the dashed lines. A surface temperature of 2000 K and a BSE composition of the  melt are assumed for both sets of calculations.}
    \label{fig:charnoz_mf_comparison}
\end{figure}

\subsection{Model validation}

Extensive validation of the LavAtmos code is described in  \cite{van_buchem_lavatmos_2023}. Here we focus on the validation of the expanded capabilities of LavAtmos 2. \cite{charnoz_effect_2023} were the first to apply a similar method to vaporisation from a melt into an atmosphere containing only H. In Figure \ref{fig:charnoz_mf_comparison} we plotted the mole fractions of H$_2$, H$_2$O, O$_2$, SiO, SiO$_2$, Fe, Na, and K in as calculated using LavAtmos 2 (solid lines) alongside the mole fractions published in \cite{charnoz_effect_2023} (dashed lines). These were calculated for a melt with a bulk silicate earth (BSE) composition \citep{palme_cosmochemical_2003} at a temperature of 2000 K. The very close match between the two outputs serves as validation of our code. However, we do observe a significant difference between the two approaches in the partial pressures calculated for SiO$_2$. LavAtmos 2 makes use of FastChem \citep{stock_fastchem_2018,stock_fastchem_2022,kitzmann_fastchem_2023} for the equilibrium gas-chemistry, while \cite{charnoz_effect_2023} use the CEA/NASA code \citep{gordon_computer_1994}. Both codes make use of the JANAF-NIST thermochemical tables \citep{chase_nist-janaf_1998} to source their thermochemical constants, however they further complement their databases from different sources and have different approaches for minimising the Gibbs free energy of the system. This is likely the reason for the differences of SiO$_2$ seen in Figure \ref{fig:charnoz_mf_comparison}.

\section{Results}\label{sec:results}

The first subsection looks at the effect of volatile atmospheres containing a single volatile element so as to isolate and compare the individual effects of the major volatile elements C, H, N, S, and P. In the second subsection of the results we show the effect of complex volatile atmospheres by testing volatile compositions that are dominated by one or two elements but contain at least a few percent of each of the other major volatile elements (see Table \ref{tab:model_compositions}), representing atmospheres that could be expected on USP with volatile atmospheres. All of these results are calculated assuming a bulk silicate earth composition (BSE, taken from  \cite{palme_cosmochemical_2003}) for the lava ocean.

\subsection{Effect of pure volatile atmospheres}

\begin{figure*}
    \centering
    \includegraphics[width=0.85\linewidth]{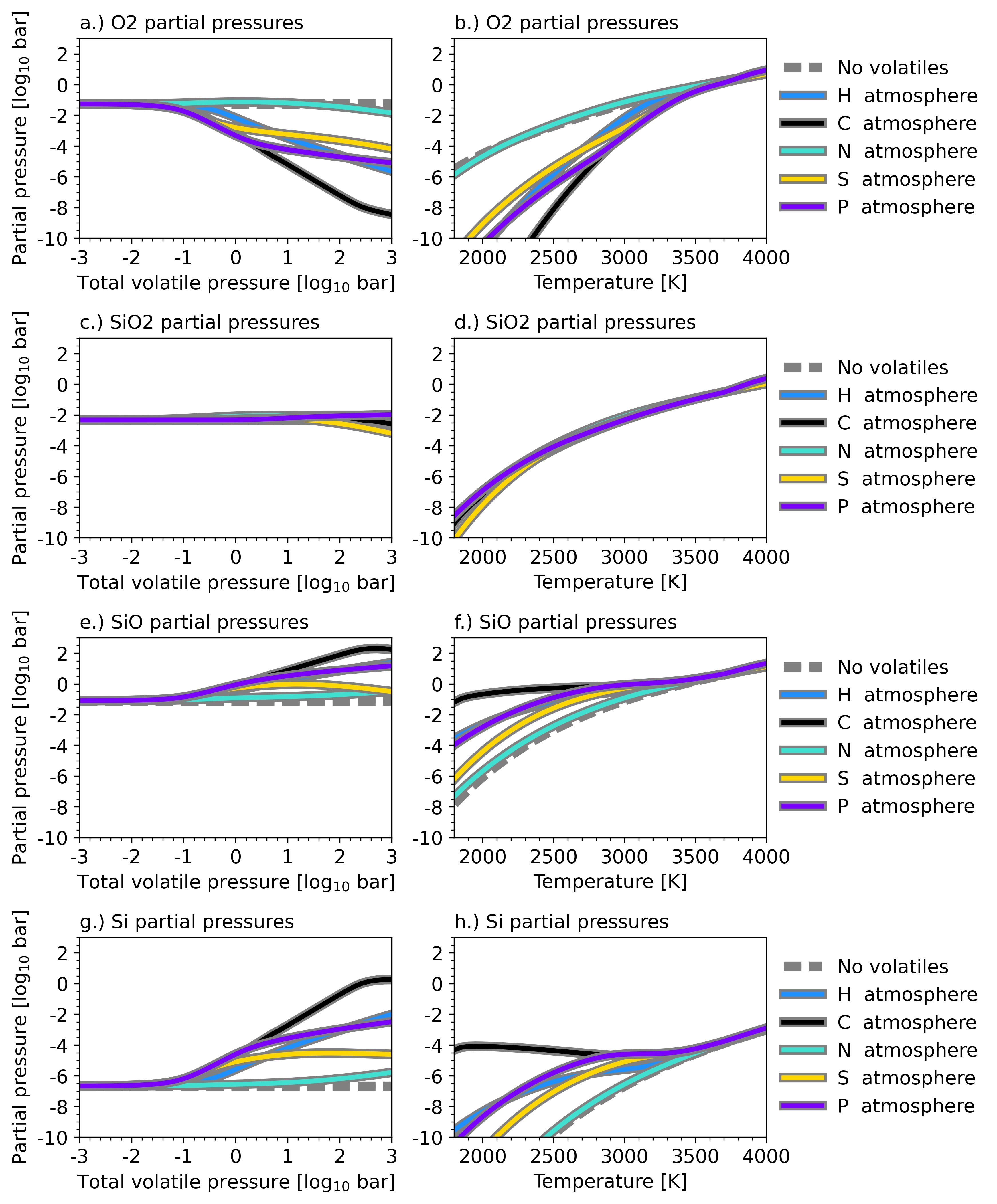}
    \caption{Effect of pure volatile atmospheres on Si species: Shown here are the partial pressures of SiO$_2$ (top row, panels \textit{a} and \textit{b}), SiO (middle row, panels \textit{c} and \textit{d}), and Si (bottom row, panels \textit{e} and \textit{f}) above a BSE lava ocean. The plots in the \textit{left} column show partial pressures as a function of total volatile pressure at a fixed temperature of 3000 K. The plots in the \textit{right} column show partial pressures as a function of temperature at a fixed total volatile pressure of 1 bar. The partial pressures of vaporised species in an atmosphere without volatiles (grey dashed line), pure H atmosphere (dark blue), pure C atmosphere (black), pure N atmosphere (light blue), pure S atmosphere (dark purple), and pure P atmosphere (light purple) above a BSE lava ocean are shown for comparison.}
    \label{fig:Pure_atmos_si}
\end{figure*}

\begin{figure*}
    \centering
    \includegraphics[width=0.68\linewidth]{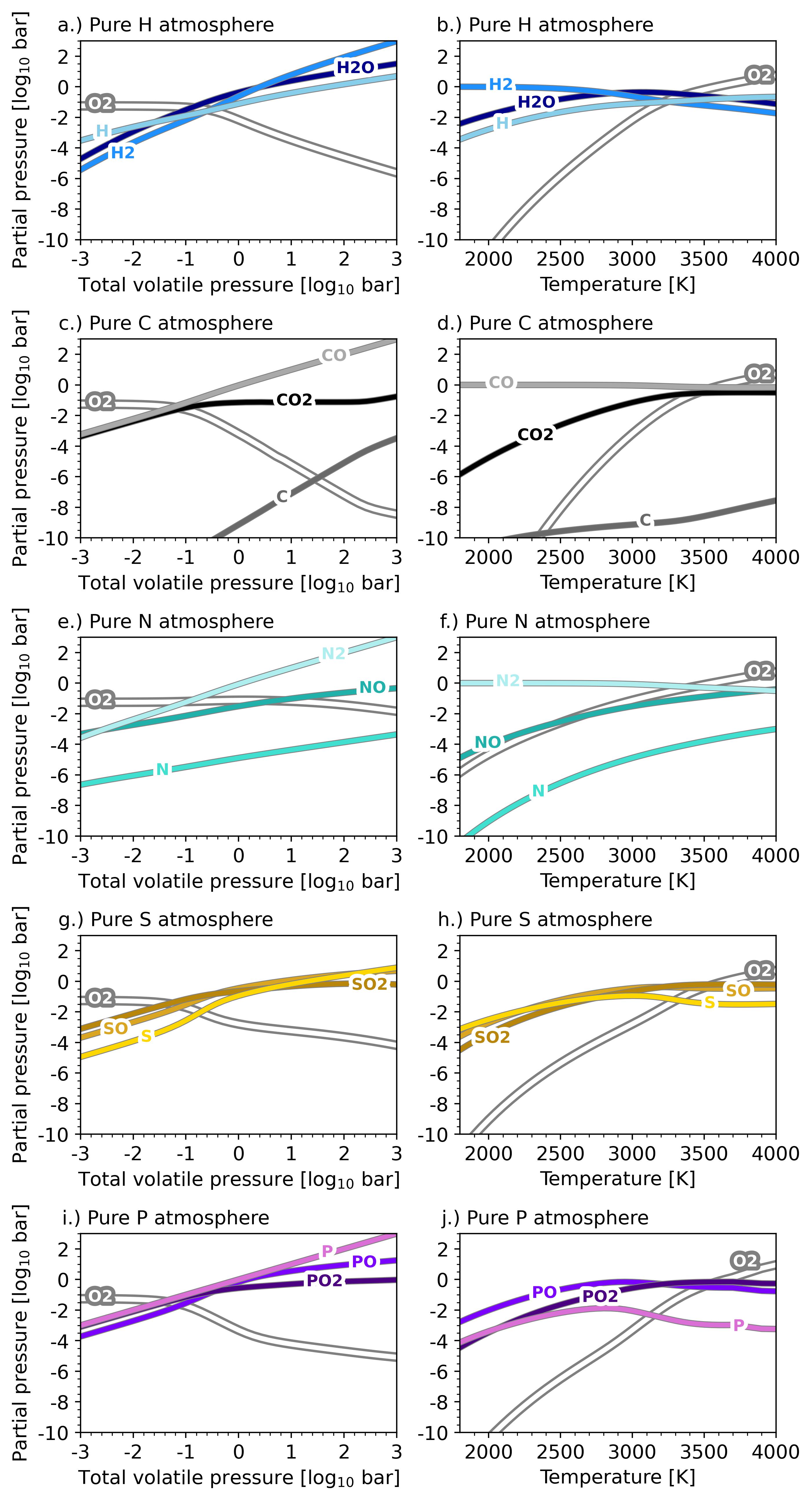}
    \caption{Behaviour of major volatile species in pure volatile atmospheres: For each of the tested pure volatile atmospheres, we show the behaviour of the major volatile species. The plots  in the \textit{left} column show partial pressures as a function of total volatile pressure at a fixed temperature of 3000 K. The plots in the \textit{right} column show partial pressures as a function of temperature at a fixed total volatile pressure of 1 bar.}
    \label{fig:Pure_atmos_vol}
\end{figure*}

\begin{figure*}
    \centering
    \includegraphics[width=0.80\linewidth]{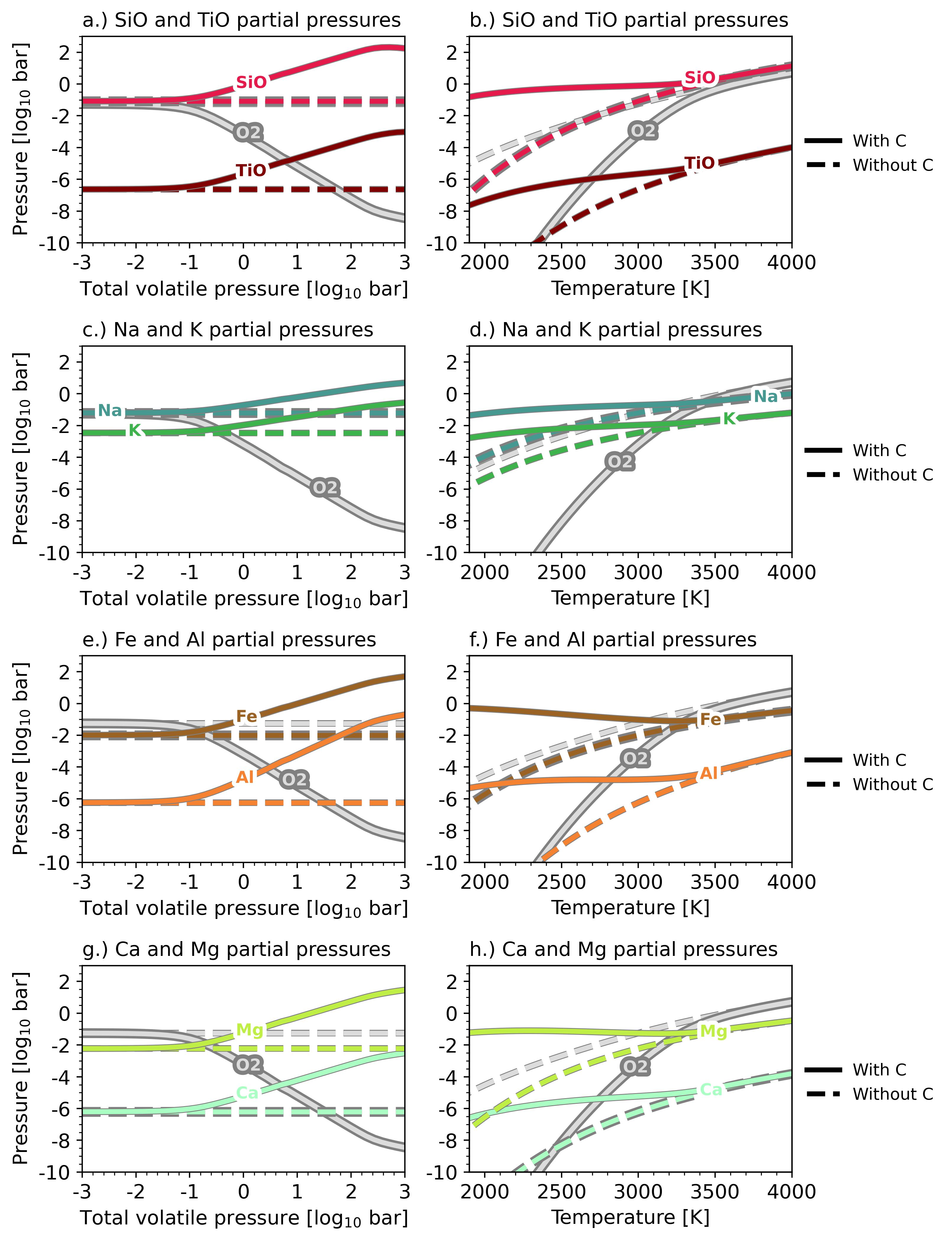}
    \caption{Effect of a pure C atmosphere: The partial pressures of selected vaporised species (O$_2$, SiO, TiO, Fe, Na, and K) in a pure C volatile atmosphere above a BSE lava ocean are shown using solid lines, while the dashed lines are used to indicate their partial pressures in volatile free atmospheres. In the bottom row (\textit{i} and \textit{j}) we also show the partial pressures of the dominant C species CO and CO$_2$. The left panels (\textit{a},\textit{c},\textit{e},\textit{g}, and \textit{i}) show the partial pressures as a function of total volatile pressure at a fixed surface temperature of 3000 K. The right panels (\textit{b},\textit{d},\textit{f},\textit{h}, and \textit{j}) show the partial pressures as a function of surface temperature at a fixed total volatile pressure of 1 bar.}
    \label{fig:vapor_spec_increase}
\end{figure*}

\begin{table}[!htbp]
\centering
\caption{Volatile components in complex atmosphere compositions above a BSE lava ocean considered in this work}
\label{tab:model_compositions}
\resizebox{\columnwidth}{!}{%
\begin{tabular}{lrrrrrc}
\multicolumn{1}{c}{\textbf{\begin{tabular}[c]{@{}c@{}}Atmosphere \\ type\end{tabular}}} & \multicolumn{1}{c}{\begin{tabular}[c]{@{}c@{}}C \\ {[}w\%{]}\end{tabular}} & \multicolumn{1}{c}{\begin{tabular}[c]{@{}c@{}}H \\ {[}w\%{]}\end{tabular}} & \multicolumn{1}{c}{\begin{tabular}[c]{@{}c@{}}N \\ {[}w\%{]}\end{tabular}} & \multicolumn{1}{c}{\begin{tabular}[c]{@{}c@{}}S \\ {[}w\%{]}\end{tabular}} & \multicolumn{1}{c}{\begin{tabular}[c]{@{}c@{}}P \\ {[}w\%{]}\end{tabular}} & \begin{tabular}[c]{@{}c@{}}O \\ {[}w\%{]}\end{tabular} \\ \hline
 & \multicolumn{1}{l}{} & \multicolumn{1}{l}{} & \multicolumn{1}{l}{} & \multicolumn{1}{l}{} & \multicolumn{1}{l}{} & \multicolumn{1}{l}{} \\
\multicolumn{7}{l}{Archetypal complex atmospheres} \\
 & \multicolumn{1}{l}{} & \multicolumn{1}{l}{} & \multicolumn{1}{l}{} & \multicolumn{1}{l}{} & \multicolumn{1}{l}{} & \multicolumn{1}{l}{} \\
\multicolumn{1}{r}{C dominated} & 80.00 & 10.00 & 9 & 0.50 & 0.50 & - \\
\multicolumn{1}{r}{H dominated} & 1.00 & 80.00 & 18.00 & 0.50 & 0.50 & - \\
\multicolumn{1}{r}{N dominated} & 1.00 & 18.00 & 80.00 & 0.50 & 0.50 & - \\
\multicolumn{1}{r}{S dominated} & 5.00 & 1.00 & 4.00 & 60.00 & 30.00 & - \\
 & \multicolumn{1}{l}{} & \multicolumn{1}{l}{} & \multicolumn{1}{l}{} & \multicolumn{1}{l}{} & \multicolumn{1}{l}{} & \multicolumn{1}{l}{} \\
\multicolumn{7}{l}{55-Cnc e best fits (Hu et al., 2024)} \\
 & \multicolumn{1}{l}{} & \multicolumn{1}{l}{} & \multicolumn{1}{l}{} & \multicolumn{1}{l}{} & \multicolumn{1}{l}{} & \multicolumn{1}{l}{} \\
\multicolumn{1}{r}{Carbon poor} & 9.09e-6 & 99.01 & 0.90 & 3.33e-6 & 6.67e-6 & \multicolumn{1}{r}{9.09e-7} \\
\multicolumn{1}{r}{Carbon rich} & 0.91 & 79.21 & 0.79 & 6.67 & 3.33 & \multicolumn{1}{r}{9.09}
\end{tabular}}
\tablefoot{The percentages are relative to the total amount of volatiles (C, H, N, S and P) added. The first four compositions, labelled "archetypal complex atmospheres", are first-order approximations of possible atmospheres dominated by one of the volatile elements. The other two compositions, labelled "55-Cnc e best fits" are volatile atmospheres that were found to be consistent with recent JWST MIRI and NIRCam observations of 55-Cnc e \citep{hu_secondary_2024}.}

\end{table}

As mentioned in section \ref{sec:methods}, we assume that all of the atmospheric oxygen comes from the vaporisation of melt. When there are no volatile elements present in the atmosphere, vaporised O$_2$ is an abundant species in the atmosphere. However, if we add the volatile elements H, C, N, S, and/or P to the atmosphere, these react with O$_2$ to form oxygenated species which reduces the O$_2$ partial pressure. This affects the vaporisation from the melt. As shown in the methods section (equation \ref{eq:PSiO}), the law of mass action dictates that the partial pressure of SiO is inversely proportional to the square root of the $\text{O}_2$ partial pressure of the atmosphere above the melt. All vaporisation reactions in which O$_2$ is released are inversely proportional to P$_{\text{O}_2}^n$ where $n$ is the amount of O$_2$ released in the reaction. So, the more O$_2$ released in the vaporisation reaction, the stronger the dependence of its partial pressure on O$_2$. Due to the formation of oxygenated species in volatile atmospheres, such as H$_2$O and CO$_2$, the O$_2$ partial pressure drops significantly when compared to non-volatile atmospheres, which increases the amounts of vaporised elements in the atmosphere.  

This effect is demonstrated in Figures \ref{fig:Pure_atmos_si}, \ref{fig:Pure_atmos_vol}, and \ref{fig:vapor_spec_increase}. In Figure \ref{fig:Pure_atmos_si} each row shows the partial pressures of a single species (O$_2$, SiO$_2$, SiO, and Si) as a function of total volatile pressure at a fixed surface temperature of 3000 K (left column) and as a function of surface temperature at a fixed total volatile pressure of 1 bar (right column). The non-volatile atmosphere case is shown using the dashed grey lines alongside the values for pure H, C, N, S, and P volatile atmospheres (where `pure' means that there is only one volatile element present in the atmosphere into which the melt is vaporising). The total atmospheric pressure of a pure H atmosphere with a volatile pressure of 1 bar is equal to the sum of 1 bar and the total pressure of vaporised species. In Figure \ref{fig:Pure_atmos_vol} we show how the major volatile species behave for each `pure' volatile atmosphere.

At 3000 K the total pressure of a pure melt vapour atmosphere is about 0.2 bar \citep{van_buchem_lavatmos_2023}. In panel \textit{a} of Figure \ref{fig:Pure_atmos_si} we see that as the total volatile pressure goes beyond 0.1 bar there is a decrease in the O$_2$ partial pressure for all volatile atmospheres. The smallest decrease is for the N atmosphere, while the largest decrease is for the C atmosphere. This is because C is much more likely to form oxygenated species (e.g. CO$_2$ and CO) than N. This is demonstrated in Figure \ref{fig:Pure_atmos_vol} when comparing panels \textit{c} and \textit{e}. In panel \textit{c} the behaviour of the major carbon species is such that as the total volatile pressure increases, the CO partial pressure grows steadily along with an initial increase in CO$_2$ and a resulting decrease in O$_2$ partial pressure. In panel \textit{e} we see that O$_2$ partial pressure decreases far less with increasing total volatile pressure due to only one of the major nitrogen species in the N atmosphere containing O (NO). From this same figure we can see that the C atmosphere is most effective at decreasing the O$_2$ partial pressure due to the fact that C is less prone to forming monoatomic species than the other four volatiles considered here.

In panel \textit{c} we see that since SiO$_2$ is not dependent on O$_2$ partial pressure in its vaporisation reaction (equation \ref{eq:PSiO2}), its partial pressure is barely affected by an increase in total volatile pressure for all atmospheres (except for the S atmosphere, see following subsection). SiO (panel \textit{e}) and Si (panel \textit{g}) partial pressures are affected, due to their dependence on O$_2$ partial pressure, and show strong increases in partial pressure with increasing total volatile pressure. The strength of the increase in the SiO and Si partial pressures is proportional to the strength in the decrease in O$_2$ partial pressures. Hence, a C atmosphere has the strongest effect, the N atmosphere the weakest, and the other three lie in between these two extremes. 

The right column of Figure \ref{fig:Pure_atmos_si} shows that a decrease in O$_2$ partial pressures with decreasing lava ocean surface temperature (panel \textit{b}) leads to an increase in SiO and Si partial pressures (panels \textit{f} and \textit{h}) relative to the volatile-free case but little to no change in SiO$_2$ partial pressure (panel \textit{d}) relative to the case in which there are no volatiles in the atmosphere above the melt. At about 3500 K, partial pressure values converge with the non-volatile atmosphere case. This is due to the fact that the pressure of vaporised species increases exponentially as a function of temperature \citep{van_buchem_lavatmos_2023}), whereas the total volatile pressure is being held constant. Hence, the partial pressures of the vaporised species overwhelm those of the volatiles. Below about 3500 K, the atmosphere is dominated by volatiles. As the temperature decreases, what little O$_2$ is being released into the atmosphere combines with the volatiles which pushes the O$_2$ partial pressure even lower, leading to very strong increases in the partial pressures of the species which are dependent on O$_2$ in their vaporisation reactions (e.g. SiO and Si). Again, the strength of this effect is strongly dependent on the propensity of the volatile element to form oxygen-bearing molecules. 

The behaviour of SiO$_2$, SiO, and Si as shown in Figure \ref{fig:Pure_atmos_si} is qualitatively representative for all species that form through vaporisation reactions from the melt. Generally, the degree of partial pressure increase relative to the non-volatile case is dictated by the moles of O$_2$ released in the vaporisation reaction. In Figure \ref{fig:vapor_spec_increase} we show this for a selection of vaporised species which are likely candidates for detection in emission and/or transmission spectra of USP planets \cite{zilinskas_observability_2022,zilinskas_observability_2023,ito_theoretical_2015} in a `pure' C atmosphere. We chose to show the effect of a C atmosphere due to it having the most pronounced effect on the partial pressure of the vaporised species. As in Figure \ref{fig:Pure_atmos_si}, the left column shows partial pressures as a function of total volatile pressure, while the right column shows the partial pressures as a function of temperature.\footnote{The full output for all these atmosphere types is available upon request from the author.} All vapour species shown in this figure are strongly affected by the presence of a volatile atmosphere. It is noteworthy to point out that at around 2000 K, Fe may become the dominant vaporised species instead of Na due to this effect. 

\subsection{Effects of a complex volatile atmosphere}

\begin{figure*}
    \centering
    \includegraphics[width=0.85\linewidth]{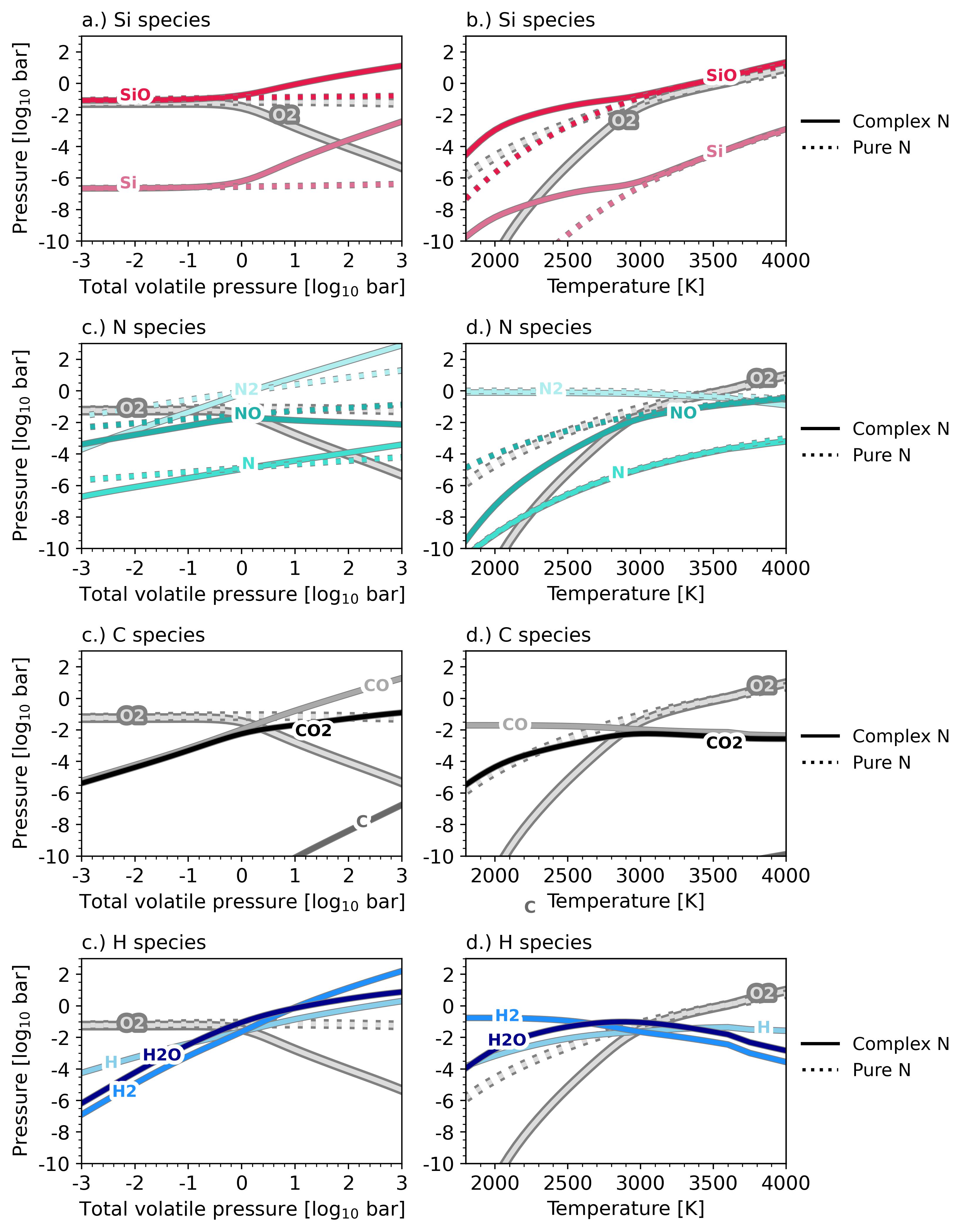}
    \caption{Chemistry of a N-dominated complex atmosphere: Shown here are the partial pressure of a selection of atmospheric species as a function of total volatile pressure at a fixed BSE lava ocean surface temperature of 3000 K in the left column (panels \textit{a},\textit{c}, \textit{e}, and \textit{g}) and temperature at a fixed total volatile pressure of 1 bar in the right column (panels \textit{b},\textit{d}, \textit{f}, and \textit{h}). The solid lines indicate partial pressures in a complex volatile atmosphere (see 'N dominated' in Table \ref{tab:model_compositions}) and the dash dotted lines a pure N atmosphere.}
    \label{fig:N_difference}
\end{figure*}

\begin{figure*}
    \centering
    \includegraphics[width=0.9\linewidth]{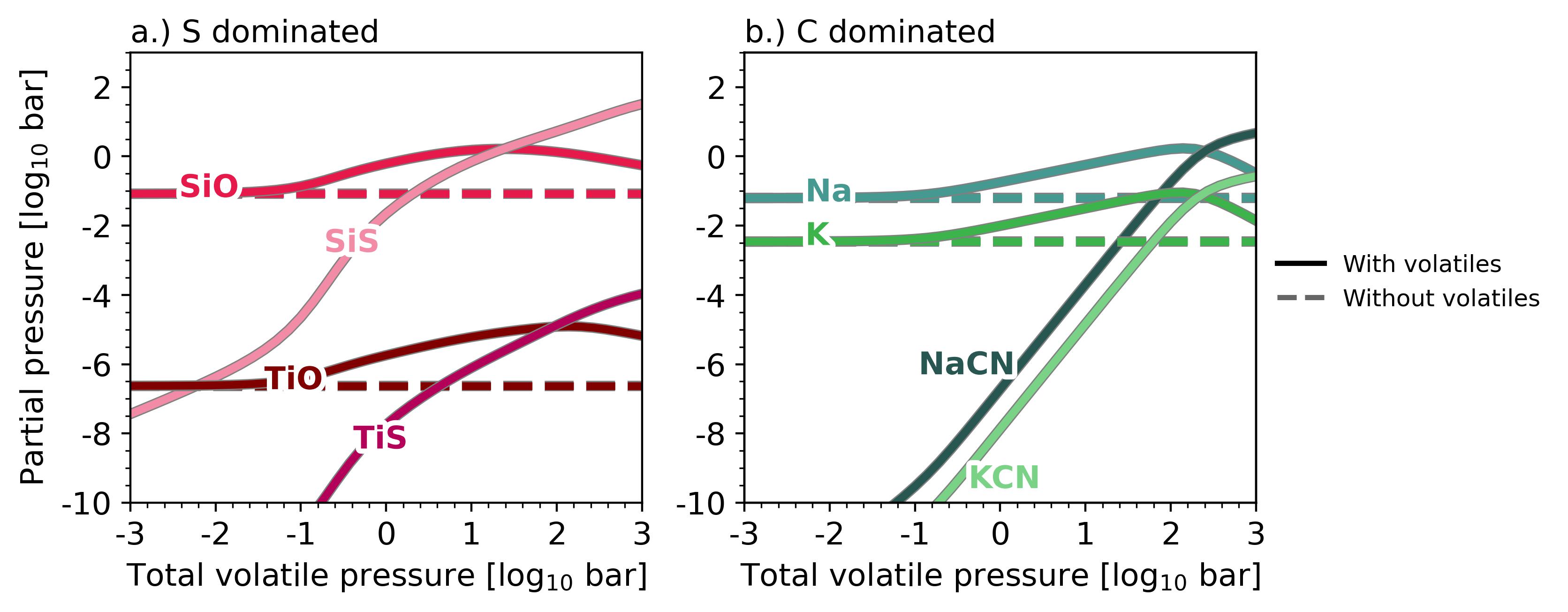}
    \caption{Behaviour of K, Na, Si, and Ti species in S- and C-dominated complex atmospheres: In both panels the calculated partial pressures of selected species are shown as a function of total volatile pressure. In the left panel (\textit{a}), a S-dominated complex atmosphere yields a decrease in SiO and TiO partial pressure due to the formation of SiS and TiS, respectively. In the right panel (\textit{b}) similar behaviour is seen in a C-dominated complex atmosphere where the partial pressures of Na and K decrease due to the formation of NaCN and KCN, respectively. Calculations for both panels were done for a fixed BSE lava ocean surface temperature of 3000 K.}
    \label{fig:Na_K_Si_Ti_variation}
\end{figure*}

To understand the effects of volatile atmospheres on vaporisation when all five major volatile elements (C, H,  N, S, P) are present in the same atmosphere, we modelled complex volatile atmospheres. It is, as of yet, not clear what the (volatile) composition of atmospheres on earth, super-earth, and sub-Neptune sized exoplanets is likely to be. The current consensus is that a wide range of different compositions is possible \citep{moses_compositional_2013,hu_photochemistry_2014,guzman-mesa_chemical_2022}. We therefore consider four different `archetypal' atmospheres that each represent a first order estimate of an atmosphere dominated by one specific element but which also contains the other four. These are not meant to represent accurate estimates of the volatile element abundances of specific HRE's, but rather to demonstrate the effect that complex volatile atmospheres have on vapour species. We tested five archetypes of atmospheres: C, H, N, and S dominated (see Table \ref{tab:model_compositions}). The C dominated atmosphere is representative of the highly enriched (x10000 solar C/H) atmosphere tested in \cite{moses_compositional_2013}, the H dominated atmosphere of a more solar-like atmosphere, the N dominated is similar to what we see on Earth and Titan, and finally the S dominated atmosphere represents what we could potentially expect on a highly volcanic planet (analogous to Io). Due to how similar the behaviour is of S and P, we decided to add a significant amount of P to the S dominated atmosphere instead of trying a separate case where P was the only dominant volatile element.

The effect of complex volatile atmospheres on the vaporised species is qualitatively the same as for the pure atmospheres (shown in Figures \ref{fig:vapor_complex_C}, \ref{fig:vapor_complex_H}, \ref{fig:vapor_complex_N}, and \ref{fig:vapor_complex_S} in the Appendix) with the dominant volatile element dictating atmospheric chemistry. 

One notable exception is the atmosphere with N as the dominant volatile, which is highlighted in Figure \ref{fig:N_difference}. In the first row, we see that the complex N dominated atmosphere leads to a significant increase in Si and SiO partial pressure at high total volatile pressures (right side of panel \textit{a}) and at relatively lower temperatures (left side of panel \textit{b}), compared to the pure N atmosphere and the case without volatiles. This indicates that N dominated atmospheres may also have significant amounts of vaporised species if some of the volatiles that are more prone to oxidation, such as C and H, are also present in the atmosphere. This is shown in the panels displaying the partial pressures of the volatile species in the atmosphere. In panels \textit{c} and \textit{d} it is shown how the complex atmosphere has a lower NO partial pressure than its pure counterpart due to the fact that O$_2$ is used up in the formation of CO, CO$_2$, and H$_2$O (panels \textit{e}, \textit{f}, \textit{g}, and \textit{h}). This leads to a lower O$_2$ partial pressure than in the pure atmosphere, together with increases in the SiO and Si partial pressures.

Expanding the number of chemical species included in LavAtmos 1 by using FastChem also enables us to study changes in partial pressures of vaporised species not only due to a change in the O$_2$ partial pressure, but also due to different species competing for the same element. Figure \ref{fig:Na_K_Si_Ti_variation} shows some examples of this for a complex S dominated atmosphere (panel \textit{a}) and a complex C dominated atmosphere (panel \textit{b}) at a fixed temperature of 3000 K. In the S dominated atmosphere, SiO and TiO initially increase as the total volatile pressure increases. SiS and TiS are being formed simultaneously, and as the total volatile pressure reaches 10 bar, these S bearing species become the most abundant Si species and the partial pressures of SiO and TiO decrease. This behaviour can also be seen in panels \textit{e} and \textit{g} of Figure \ref{fig:Pure_atmos_si} for the pure S atmosphere. Figure \ref{fig:Na_K_Si_Ti_variation}\textit{b} shows a feature that is unique to a complex volatile atmosphere that includes both C and N. Here, Na and K initially see an increase in partial pressure with increasing total volatile pressure, but as the total volatile pressure reaches 100 bar, NaCN and KCN become the dominant Na and K bearing species. This example serves to illustrate the importance of including a wide range of different chemical species in order to fully understand the behaviour of the major species over different temperatures and pressures. 

\section{Implications for a 55-Cnc e-like atmosphere}\label{sec:55-Cnc-e}

To assess the implications of our results, we modelled the composition at the surface-atmosphere boundary of a planet analogous to 55-Cnc e. Recent JWST observations of the emission spectrum of 55-Cnc e gathered using MIRI-LRS and NIRCam were published by \cite{hu_secondary_2024}. The observed spectrum does not match that of a thin vaporised-rock atmosphere and instead matches better with the spectrum of a $\simeq$100 bar volatile atmosphere that would facilitate more efficient heat-redistribution. Furthermore, absorption features found in the NIRCam spectrum are consistent with CO and CO$_2$ features. This points to 55-Cnc e supporting a volatile atmosphere. These main conclusions are based on the differences between two end-member models: a volatile-free atmosphere and a volatile-bearing but rock-vapour-free atmosphere. \cite{hu_secondary_2024} also determined the range of physically plausible atmospheres on the planet using self-consistent atmospheric models that contained both rock-vapour and volatiles.

Of these self-consistent atmospheric models, the majority were run assuming only volatile species in the atmosphere without a lava ocean feeding rock-vapour species into it. These models indicate that the best fitting atmosphere is one containing abundant CO$_2$ with a surface temperature of about 2500 K. Some models included species from both lava vaporisation and volatile species, using the approach described in  \cite{zilinskas_observability_2023}, which is similar to the approach of \cite{piette_rocky_2023}. The abundances of the vaporised species were calculated without taking the influence of volatile species on vaporisation into account. These abundances were then added to the total abundances of the atmosphere. We refer to this as the `volatile-free vaporisation' method (which is analogous to the 'sum' method used in the work of \cite{falco_hydrogenated_2024}). This approach gave an overall worse fit to the JWST data than the volatile-only scenario, but did indicate that a reasonable fit could be made with a higher surface temperature of a bit less than 3000 K.

Finally, \cite{hu_secondary_2024} tested a third type of model in which the vaporisation calculations used an early version of the code presented in this paper. This led to a large increase in the abundance of the vaporised species in the atmosphere, causing a strong inversion of the temperature-pressure profile and yielding a spectrum with strong emission features. This shows that if the conflicting data points observed around 4.5 micron by previous observations of 55-Cnc e with Spitzer \citep{demory_map_2016,demory_variability_2016} are indeed physical, this could potentially be explained by an influx in vaporised species into the atmosphere of 55-Cnc e. This model also provides a potential fit with a significantly lower surface temperature of only 2000 K.

From these models, one can conclude (a) that a wide range of different surface temperatures could potentially explain the observed emission spectrum of 55-Cnc e and (b) that abundances of rock-vapour species in the atmosphere have important implications for the thermal structure of the atmosphere and resulting emission/absorption features of the spectrum. Here we assess the extent to which the \cite{hu_secondary_2024} results are affected when using LavAtmos 2.0.  We decided to see how two of the best fitting volatile atmosphere compositions from \cite{hu_secondary_2024} affect vaporisation when coupled to a lava-ocean.

We selected two model atmospheric volatile compositions with very different C abundances (see Table \ref{tab:model_compositions}) which both show a good fit to the observed spectra of 55-Cnc e. As we have seen in the previous section, the presence of C in the atmosphere has a strong effect on the vaporisation of melt species. We labelled the compositions based on their C abundances, with one named carbon poor (C w\% $= 9.09$e-6) and the other named carbon rich (C w\% $= 0.91$). 

\begin{figure*}
    \centering
    \includegraphics[width=0.75\linewidth]{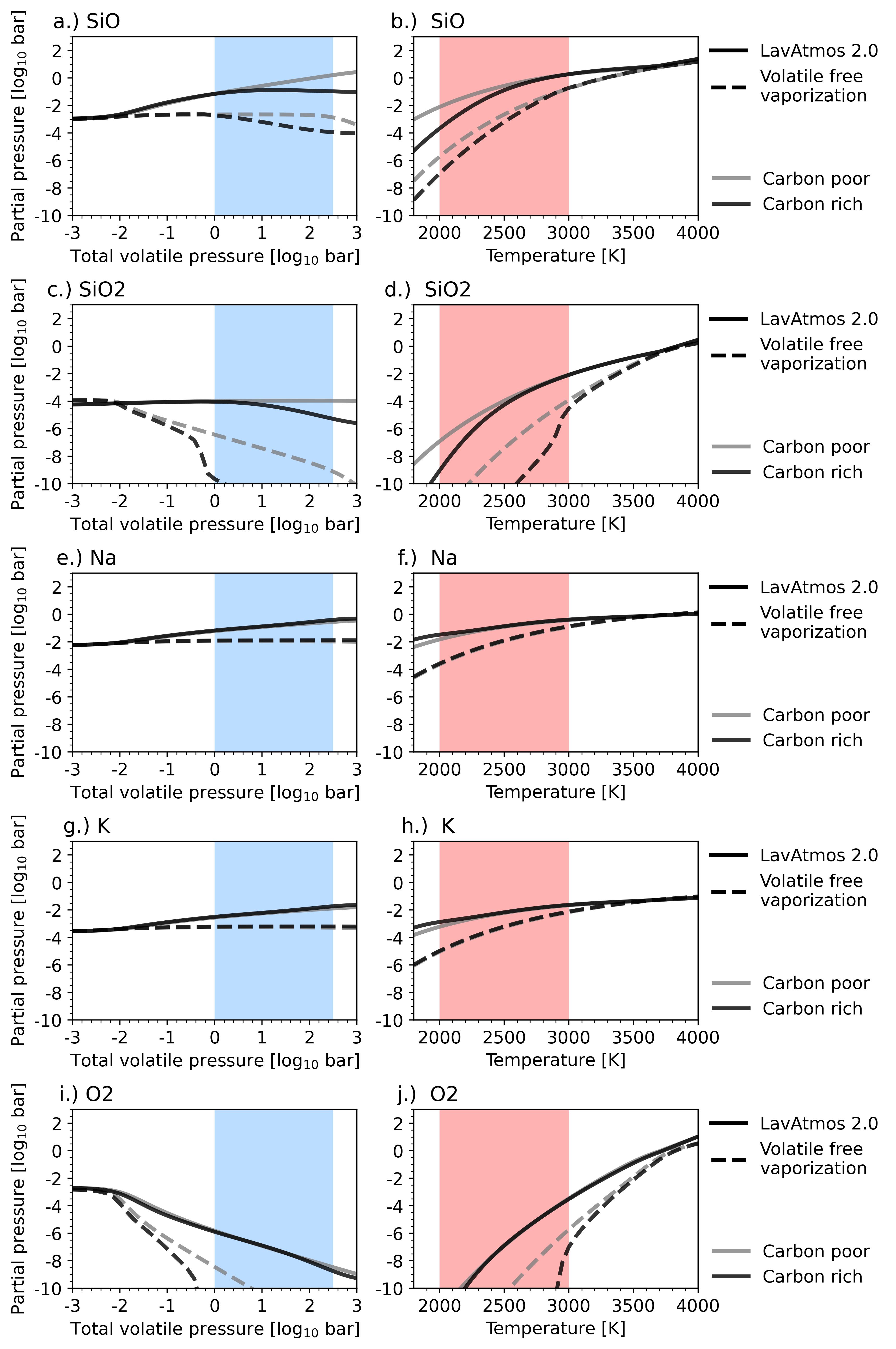}
    \caption{Computed partial pressures of vaporised species in a 55-Cnc e like atmosphere: Shown here are the partial pressures of SiO (panels \textit{a} and \textit{b}),  SiO$_2$ (panels \textit{c} and \textit{d}), Na  (panels \textit{e} and \textit{f}), K (panels \textit{g} and \textit{h}), and O$_2$ (panels \textit{i} and \textit{j}). The solid lines indicate the partial pressures calculated using LavAtmos 2.0 (this study), and the dashed lines show the partial pressures calculated when using the volatile-free vaporisation approach (LavAtmos 1). Two volatile atmosphere compositions are tested (see Table \ref{tab:model_compositions}) of which one is carbon poor ($\simeq$0.01 w\% C, grey) and the other carbon rich ($\simeq$10\%, black). The left-hand panel shows partial pressures plotted as a function of the total volatile pressure of the atmosphere at a fixed temperature of 2500, and the right-hand panel as a function of surface temperature at a fixed total volatile pressure of 10 bar. The blue area indicates the estimated surface pressure range of 55-Cnc e and the red area indicates the estimated surface temperature range of 55-Cnc e \citep{hu_secondary_2024}.}
    \label{fig:55-cnce_models_vapo}
\end{figure*}

\begin{figure*}
    \centering
    \includegraphics[width=0.8\linewidth]{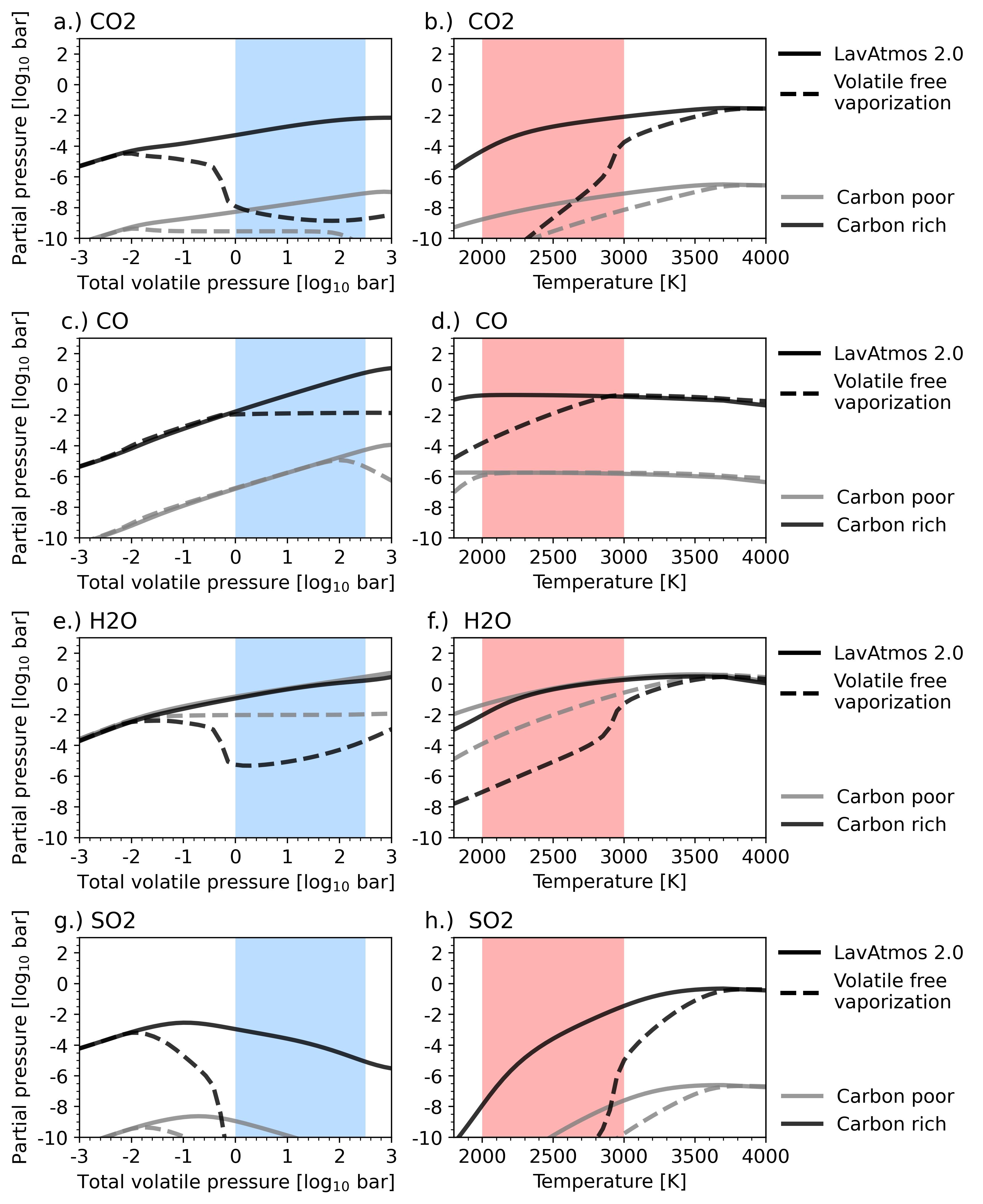}
    \caption{Computed partial pressures of volatile species in a 55-Cnc e like atmosphere: Shown here are the partial pressures of CO (panels \textit{a} and \textit{b}) CO$_2$ (panels \textit{c} and \textit{d}), H$_2$O (panels \textit{e} and \textit{f}), SO$_2$ (panels \textit{g} and \textit{h}). The solid lines indicate the partial pressures calculated using LavAtmos 2.0 (this study) and the dashed lines show the partial pressures calculated when using the volatile-free vaporisation approach. Two volatile atmosphere compositions are tested (see Table \ref{tab:model_compositions}) of which one is carbon poor ($\simeq$0.01 w\% C, grey) and the other carbon rich ($\simeq$10\%, black). The left-hand panel shows partial pressures plotted as a function of the total volatile pressure of the atmosphere at a fixed temperature of 2500, and the right-hand panel as a function of surface temperature at a fixed total volatile pressure of 10 bar. The blue area indicates the estimated surface pressure range of 55-Cnc e and the red area indicates the estimated surface temperature range of 55-Cnc e \citep{hu_secondary_2024}.}
    \label{fig:55-cnce_models_vol}
\end{figure*}

For these two volatile compositions we plotted the partial pressures, calculated using both LavAtmos and the volatile-free vaporisation method, of O$_2$ along with important vaporised species (SiO, SiO$_2$, Na, and K) in Figure \ref{fig:55-cnce_models_vapo} and the partial pressures of volatile species with strong spectroscopic signatures (CO, CO$_2$, H$_2$O, and SO$_2$) in Figure \ref{fig:55-cnce_models_vol}. It is important to note that the partial pressures of the volatile-free vaporisation method are calculated by first deriving the partial pressures of the vapor species while assuming that no volatile species are present. From these partial pressures, the elemental abundance of the vaporised elements (O, Si, Mg, Al, Ti, Fe, Ca, Na, and K) are derived, which are then combined with the elemental abundances of the volatile elements (C, H, N, S, and P) and provided as input to FastChem 3 \citep{stock_fastchem_2018,kitzmann_fastchem_2023}, which calculates the partial pressures of the chemical species at the desired TP point.

Similar to the previous section, we show output for a fixed temperature and varying the total volatile pressure in the panels in the left column and a fixed total volatile pressure but varying temperature in the right side column. We use a fixed surface temperature of 2500 K and a fixed total volatile pressure of 10 bar. These values fall in the middle of the range of total volatile pressures and surface temperatures (indicated using the blue and red shaded regions in Figures \ref{fig:55-cnce_models_vapo} and \ref{fig:55-cnce_models_vol}) that could be realistically expected based on the tested self-consistent models in \cite{hu_secondary_2024} as summarised above.

In this section, we first draw attention to the differences in composition that the two approaches (`volatile-free vaporisation' and the method presented in this paper) of calculating the composition of melt-vapour in a volatile atmosphere can give. We then look at how the presence or absence of a lava-ocean at the base of a volatile atmosphere could be reflected in its composition.

\begin{figure*}
    \centering
    \includegraphics[width=0.9\linewidth]{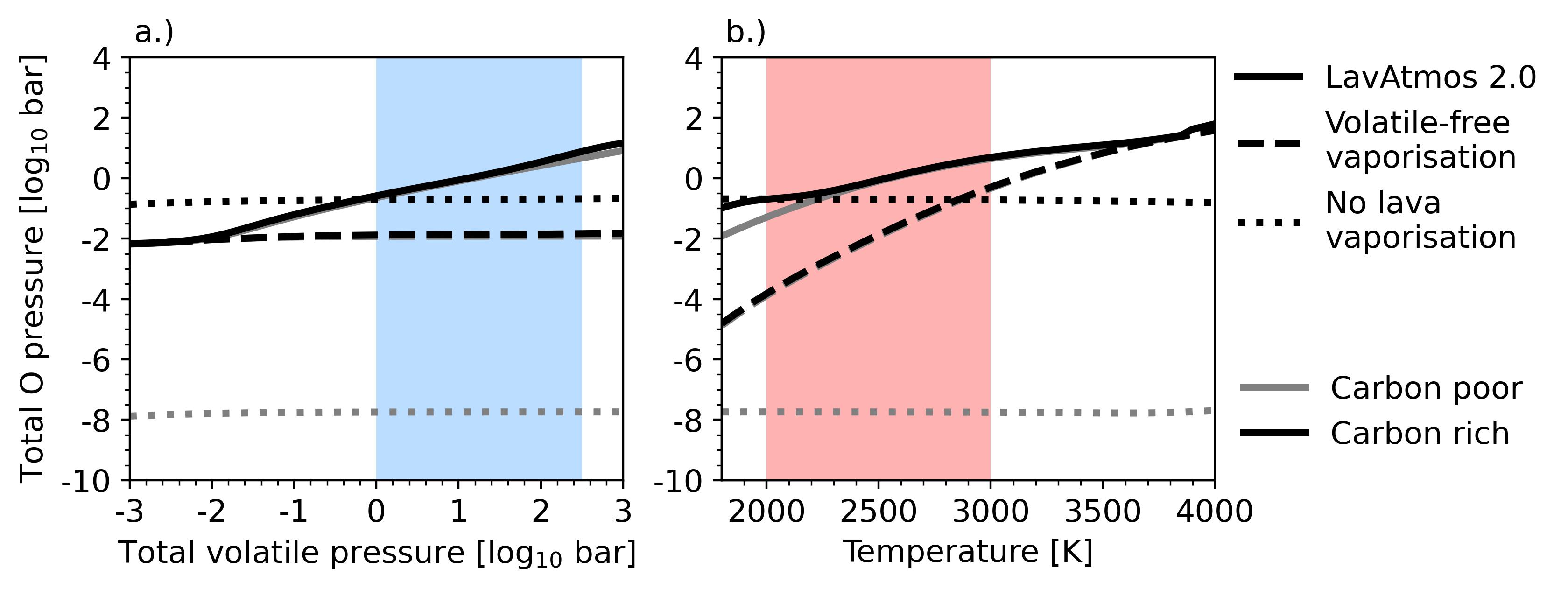}
    \caption{Total O pressure in a 55-Cnc e like atmosphere: Shown here are the total O pressure of two different volatile atmosphere compositions - carbon poor (\textit{grey}) and carbon rich (\textit{black}) as a function of total volatile pressure (panel \textit{a}) at a fixed temperature of 2500 K and as a function of temperature at a fixed total volatile pressure of 10 bar (panel \textit{b}). The total O pressures indicated using the solid lines are calculated using the method outline in this paper, those indicated using the dashed line are calculated using the volatile-free vaporisation method, and the dotted line indicates the total O pressure when no lava ocean vaporisation is assumed. The blue area indicates the estimated surface pressure range of 55-Cnc e and the red area indicates the estimated surface temperature range of 55-Cnc e \citep{hu_secondary_2024}.}
    \label{fig:55-cnce_O_ab}
\end{figure*}

\subsection{Vaporised partial pressure comparison between LavAtmos 2.0 and the volatile-free vaporisation method}\label{sec:lavatmos2vsvolfree}

Looking first at the total volatile pressure dependent plots in the left-side column of Figure \ref{fig:55-cnce_models_vapo}, we see how at low total volatile pressure the partial pressures calculated by LavAtmos 2.0 and the volatile-free vaporisation method are in agreement. At these conditions, the abundances of volatile species are too low to have a significant effect on the vaporisation. As we move to the right in these plots, the partial pressures calculated by the two methods start to diverge around 1e-2 bar for all species. This is due to the fact that at the surface temperature at which the volatile pressure dependent calculations were made (2500 K) the total vapour pressure above a melt of BSE composition in a volatile-free atmosphere is around 1e-2 bar \citep{visscher_chemistry_2013,ito_theoretical_2015,van_buchem_lavatmos_2023}. As the total volatile pressure nears and then surpasses this point, it starts dominating the atmosphere.

As we have seen in section \ref{sec:results}, the presence of volatile elements in the atmosphere leads to a drop in O$_2$ partial pressure, which leads to an increase in the vaporisation of species for which O$_2$ is released in their vaporisation reaction. This effect occurs here as well, as SiO, Na, and K all exhibit increasing partial pressures relative to the volatile-free vaporisation case once the total volatile pressures go beyond 1e-2 bar. 

This may appear to be in contradiction to what we see in the bottom two panels (\textit{i} and \textit{j}) where the O$_2$ partial pressure of the volatile-free atmosphere is lower than that of LavAtmos 2. However, this is a product of how the volatile-free vaporisation approach works. As explained at the start of section \ref{sec:55-Cnc-e}, the volatile-free vaporisation approach first calculates the abundance of the vaporised elements and only then combines them with the volatile elements to calculate the chemical composition of the atmosphere. Hence, during the vaporisation calculations, the O$_2$ found using the volatile-free vaporisation method is indeed greater than for LavAtmos 2 (similar to what we see in Figure \ref{fig:vapor_spec_increase}). It is once these partial pressures are converted to abundances, combined with the volatile elements, and the partial pressures are re-calculated, that the resulting O$_2$ partial pressure is decreased due to the formation of oxygenated species such as H$_2$O and CO$_2$.

This indirectly illustrates another distinction between the volatile-free vaporisation approach and LavAtmos 2, which is that including volatile species in the vaporisation reactions leads to an increase in the amount of O in the atmosphere. This is shown explicitly in Figure \ref{fig:55-cnce_O_ab}, where we plot the total O pressure for all three models as a function of total volatile pressure (panel \textit{a}) and temperature (panel \textit{b}). In both panels the O abundance of the atmosphere without lava vaporisation is fixed - as expected since the O abundance is set a-priori (see Table \ref{tab:model_compositions}). For the volatile-free vaporisation case, the O abundance is fixed in the total volatile pressure plot (\textit{a}) because the presence of the volatile species does not affect the vaporisation - which is also why the same O abundance is found for both volatile compositions. LavAtmos 2 does see an increase in O abundance with increasing total volatile pressure of up to three orders of magnitude with respect to the volatile free case. In panel \textit{b} in Figure \ref{fig:55-cnce_O_ab}, the total O pressure of the volatile-free vaporisation method increases with temperature, coinciding with the increase in vaporisation with higher temperatures. Here we also see that LavAtmos 2 gives a greater total O pressure due to the presence of the volatile species.

It may feel counter-intuitive that the inclusion of volatile species in the vaporisation reaction leads to an increase in O elemental abundance in the atmosphere, while we have been discussing how a drop in O$_2$ partial pressure is what is causing an increase in vaporisation partial pressures. However, this makes sense if one considers that a drop in O$_2$ partial pressure is due to the formation of oxygenated species with volatile elements. As such, the O is not being removed from the atmosphere. Simultaneously, the increase in vaporisation of melt oxides due to a drop in O$_2$ partial pressure leads to an increase in the amount of O being released into the atmosphere.

The difference in O abundance between the two approaches is also reflected in the decrease of the partial pressure of O bearing vapour species such as SiO and SiO$_2$ at high total volatile pressures (panels \textit{a} and \textit{c}). As the total volatile pressure increases, causing a drop in O$_2$ partial pressure, SiO$_2$ partial pressure and to a lesser extend SiO partial pressure decrease significantly in the volatile-free vaporisation approach. The increased O abundance seen in LavAtmos 2 entails that this effect is far less pronounced, affecting the carbon rich atmosphere more so than the carbon poor atmosphere due to the propensity of C to form oxidized species.

In the right-side column in Figure \ref{fig:55-cnce_models_vapo}, we see that at high temperatures the partial pressures calculated by LavAtmos 2.0 and the volatile-free vaporisation method generally have overlapping values. As temperature decreases, they start to diverge around 3600 K. This coincides with the temperature at which the total vaporised pressure in a volatile-free atmosphere is $\simeq10$ bar \citep{visscher_chemistry_2013,ito_theoretical_2015,van_buchem_lavatmos_2023}, which is the fixed total volatile pressure at which these plots were made. As temperature decreases further, the volatile atmosphere is dominant over the rock-vapour. As we decrease further in temperature, all partial pressures drop, but those calculated using LavAtmos 2.O do not drop as quickly due to the presence of the volatiles lowering the O$_2$ partial pressure. The drop in partial pressures of SiO and SiO$_2$ of the volatile-free vaporisation method are accelerated further due to the volatile species competing with Si for O.

With the exception of vapour atmosphere dominated regimes (at low total volatile pressure and high surface temperature), we see that the volatile-free vaporisation method systematically leads to an underestimation of the partial pressures relative to LavAtmos 2.0. The most significant effects are seen for SiO and SiO$_2$. Within the total volatile pressure and temperature ranges (indicated in red and blue respectively) that could be expected on 55-Cnc e \citep{hu_secondary_2024}, the differences in partial pressure between these two approaches are of one to several orders of magnitude. This could potentially mean that the spectral emission features of SiO and SiO$_2$ may not be as suppressed by volatile atmospheres as currently thought \citep{zilinskas_observability_2023, piette_rocky_2023}. The differences between the two methods that we see in the partial pressures of Na and K could be of relevance for the thermal structures of atmospheres, which can be heavily influenced by an increase in short-wave absorbers such as Na and K. These findings align with the work by \cite{falco_hydrogenated_2024}, which suggests that hydrogen-rich atmospheres may contain sufficient abundances of vaporised species for a thermal inversion to occur.

\begin{figure*}
    \centering
    \includegraphics[width=0.9\linewidth]{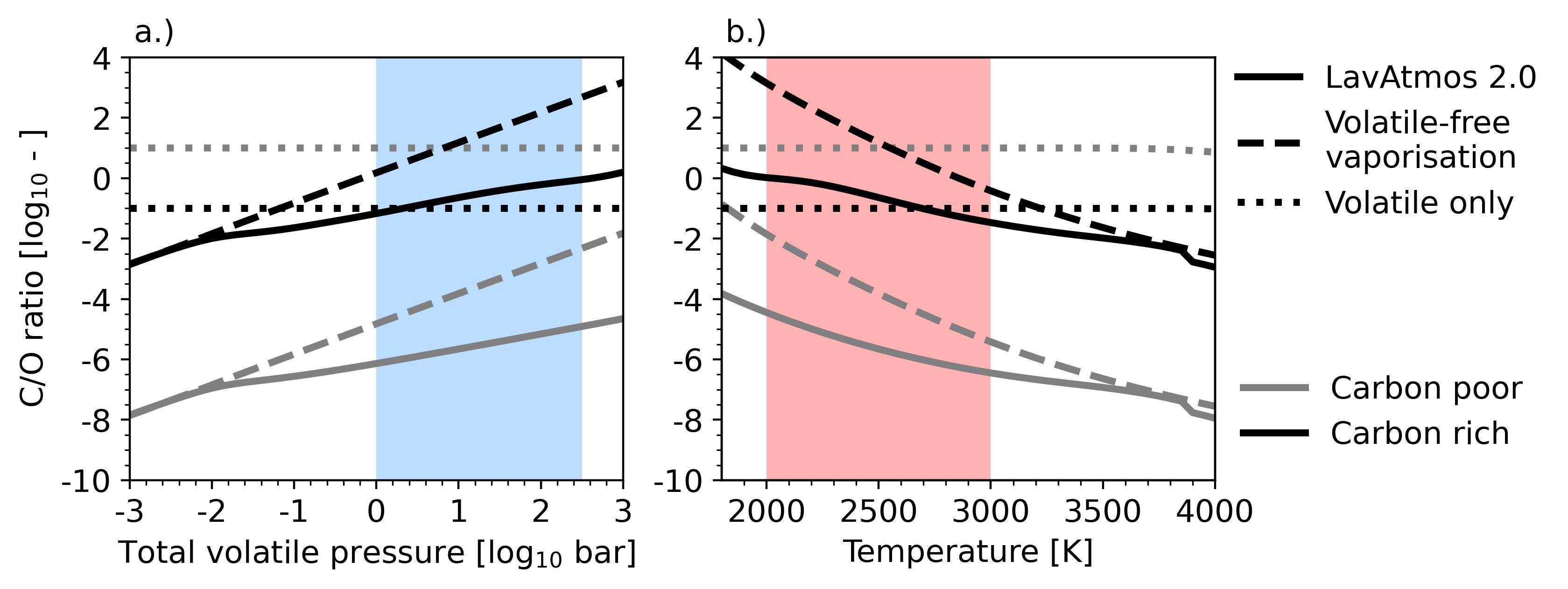}
    \caption{C/O ratio in a 55-Cnc e like atmosphere: Shown here are the C/O ratios of two different volatile atmosphere compositions - carbon poor (\textit{grey}) and carbon rich (\textit{black}) as a function of total volatile pressure (panel \textit{a}) at a fixed temperature of 2500 K and as a function of temperature at a fixed total volatile pressure of 10 bar (panel \textit{b}). C/O ratios indicated using the solid lines are calculated using the method outline in this paper, those indicated using the dashed line are calculated using the volatile-free vaporisation method, and the dotted line indicates (constant) C/O ratio when no lava ocean vaporisation is assumed. The blue area indicates the estimated surface pressure range of 55-Cnc e and the red area indicates the estimated surface temperature range of 55-Cnc e \citep{hu_secondary_2024}.}
    \label{fig:55-cnce_CO_ratio}
\end{figure*}

\subsection{The effect of lava oceans on atmospheric volatile species}\label{sec:effect_lava_ocean}

In Figure \ref{fig:55-cnce_models_vol} the change in partial pressure of spectroscopically important volatile species is shown for LavAtmos 2.0 and the volatile-free vaporisation method. Similarly to what we saw in section \ref{sec:lavatmos2vsvolfree}, at low total volatile pressure and high temperature the system is dominated by the vaporised species and the two methods have very similar outputs. However, as we move into the volatile dominated regimes, beyond 1e-2 bar in the total volatile pressure dependent plots and below 3600 K in the temperature dependent plots, we again start seeing large differences between the two approaches.

The differences we see for the volatile species can all be attributed to the fact that in LavAtmos 2.0 the presence of volatiles leads to an increase in the total O elemental abundance with respect to the the volatile-free vaporisation method as seen in the previous section and shown in Figure \ref{fig:55-cnce_O_ab}. Hence, as the volatile contents of the atmosphere increase, far fewer oxidized species can be formed when compared to the partial pressure from LavAtmos 2.0. In the left column of Figure \ref{fig:55-cnce_models_vol}, we can see that this effect is most pronounced in the fully oxidized species CO$_2$, H$_2$O, and SO$_2$. As the total volatile pressure increases beyond 1e-2 bar into the volatile dominated regime, the partial pressures of these species as calculated using the volatile-free vaporisation method decrease significantly while we see a gradual increase for LavAtmos 2.0. At around 1 bar, we see a very strong drop in CO$_2$ partial pressure for the carbon rich atmosphere. This can be attributed to the total elemental C abundance (about 1\% of the total volatile atmosphere composition as per Table \ref{tab:model_compositions}) reaching a point where it dominates over the O elemental abundance in the system. This is also reflected in the CO partial pressure; once the total volatile pressure passes 1 bar, there is too little O in the volatile-free vaporisation model to produce additional CO and the partial pressure remains constant.

Looking at the temperature dependent plots in the right-hand side column of Figure \ref{fig:55-cnce_models_vol}, we see that in the volatile dominated regime (below $\simeq$3600 K) the fully oxidised species show a strong decrease in partial pressure. This is due to the size of the vapour atmosphere being strongly dependent on temperature. As the amount of vapour in the atmosphere decreases, so does the total O elemental abundance (see panel \textit{b} in Figure \ref{fig:55-cnce_O_ab}). As expected, this decrease is much stronger in the volatile-free atmosphere than for LavAtmos 2.0 thanks to the increased vaporisation due to the presence of volatile species. For CO we see a slight increase in partial pressure with decreasing temperature; this is due to a lower O$_2$ environment favouring CO formation over CO$_2$. At 2900 K we then see that the volatile-free vaporisation method diverges from LavAtmos 2.0 (going from high to low temperatures) as there is no longer sufficient O being vaporised. The point at which this occurs is dependent on the total volatile pressure of the atmosphere. For higher total volatile pressures, a higher O abundance is required to sate the formation of oxidised species, while the inverse is true for lower total volatiles pressures - moving this point of divergence/convergence between the models up and down in temperature accordingly.

Overall, when in a volatile pressure and temperature regime where the volatile atmosphere is dominant, LavAtmos 2.0 leads to volatile species in the atmosphere being more oxidised than when using the volatile-free vaporisation method. This leads to the two methods having large differences in partial pressures of volatile species when compared to each other. Especially for the oxidised species, we see difference of more than six orders of magnitude within the total volatile pressure and temperature ranges (indicated in blue and red respectively) expected on 55-Cnc e \citep{hu_secondary_2024}. This is especially significant due to the fact that these are all species that have strong opacities within both the optical and the infrared and could therefore have large implications for the emission spectra of these atmospheres.

In order to have a more general view of the differences between the two approaches, we plotted the C/O ratios of each tested case in Figure \ref{fig:55-cnce_CO_ratio}. As one would expect, we again see a comparable trend to what we saw in Figures \ref{fig:55-cnce_models_vapo} and \ref{fig:55-cnce_models_vol} where the output of LavAtmos 2.0 and the volatile-free vaporisation method coincide within the vapour dominated regimes at low total volatile pressure and high surface temperature. The divergence happens once we enter the volatile dominated regimes (above about 1e-2 bar and below about 3600 K). In both the total volatile pressure and temperature plots, the increased O abundance seen in LavAtmos 2.0 output in the presence of volatile leads to a lower C/O ratio than the volatile-free vaporisation method. The volatile-free vaporisation method has a fixed atmospheric elemental O abundance for a given temperature, hence why we see a linear increase in the C/O ratio in the total volatile pressure dependent plot (panel \textit{a}). Again, we see that within the range of expected possible pressures and temperatures on 55-Cnc e (blue and red regions respectively) there are significant differences between the two approaches of several orders of magnitude.  

Figure \ref{fig:55-cnce_CO_ratio} also includes the fixed C/O abundances of the two hypothetical 55-Cnc e atmosphere compositions (see Table \ref{tab:model_compositions}) so as to represent the case where no lava vaporisation is taken into account, hence it is labelled as `volatile-only'. We would like to point out that one needs to be careful with the interpretation of this plot, due to the fact that these are two fundamentally different approaches. In the LavAtmos 2.0 and the volatile-free vaporisation model, it is assumed that all of the oxygen in the atmosphere comes from the vaporisation of the melt species in the lava ocean, whereas for the volatile only we assume the O composition given in Table \ref{tab:model_compositions}, which is the O abundance found to fit for these atmospheres in \cite{hu_photochemistry_2014}. Nonetheless, we consider it a worthwhile exercise to compare the calculated C/O ratios using the vaporisation methods to the fixed C/O found to give a good fit to the 55-Cnc e data, as it allows us to see how the C/O ratio varies as a function of total volatile pressure and temperature relative to values that are consistent with observations (the large majority of the tested models had C/O ratio's within the range of these two selected models).

In panel \textit{a} we see that the C/O ratio calculated by LavAtmos 2.0 for a carbon rich atmosphere matches the fixed C/O ratio of the volatile-only atmosphere well between total volatile pressures of about 1 to 10 bar. In panel \textit{c}, we see a similar match around about 2700 K. Both within the expected possible pressure and temperature ranges of 55-Cnc e. For the volatile-free vaporisation case, the C/O ratio crosses the volatile-only lines well outside of the possible pressure and temperature range at around 7e-2 bar and 3250 K. In the case of the carbon poor atmospheres, neither model is able to come even close to the C/O ratio of the volatile-only atmosphere.

If 55-Cnc e does indeed have a lava ocean at the base of its atmosphere that is also large enough to buffer the O$_2$ partial pressure of the system, then a carbon rich atmosphere with the composition given in Table \ref{tab:model_compositions} that is in equilibrium with a lava ocean, could give a C/O ratio that matches the C/O ratio of a volatile only atmosphere that is found to give a good fit to the emission spectrum. This is not the case for the carbon poor atmosphere.

We wish to stress that these outcomes are based on all of the above mentioned assumptions, along with a number of important caveats discussed in the following section. We therefore do not think that these findings provide sufficiently strong evidence to draw firm conclusions about the possible composition of the atmosphere of 55-Cnc e. However, this analysis does show that having a lava ocean in direct contact with a volatile atmosphere can have significant effects on the atmospheric oxidation state and needs to be considered in the future for a proper interpretation of the observational data of potential lava planets.

\section{Discussion}\label{sec:discussion}

\subsection{Effect of water on melt activities}\label{sec:discussion_water_effect}

\begin{figure}
    \centering
    \includegraphics[width=\linewidth]{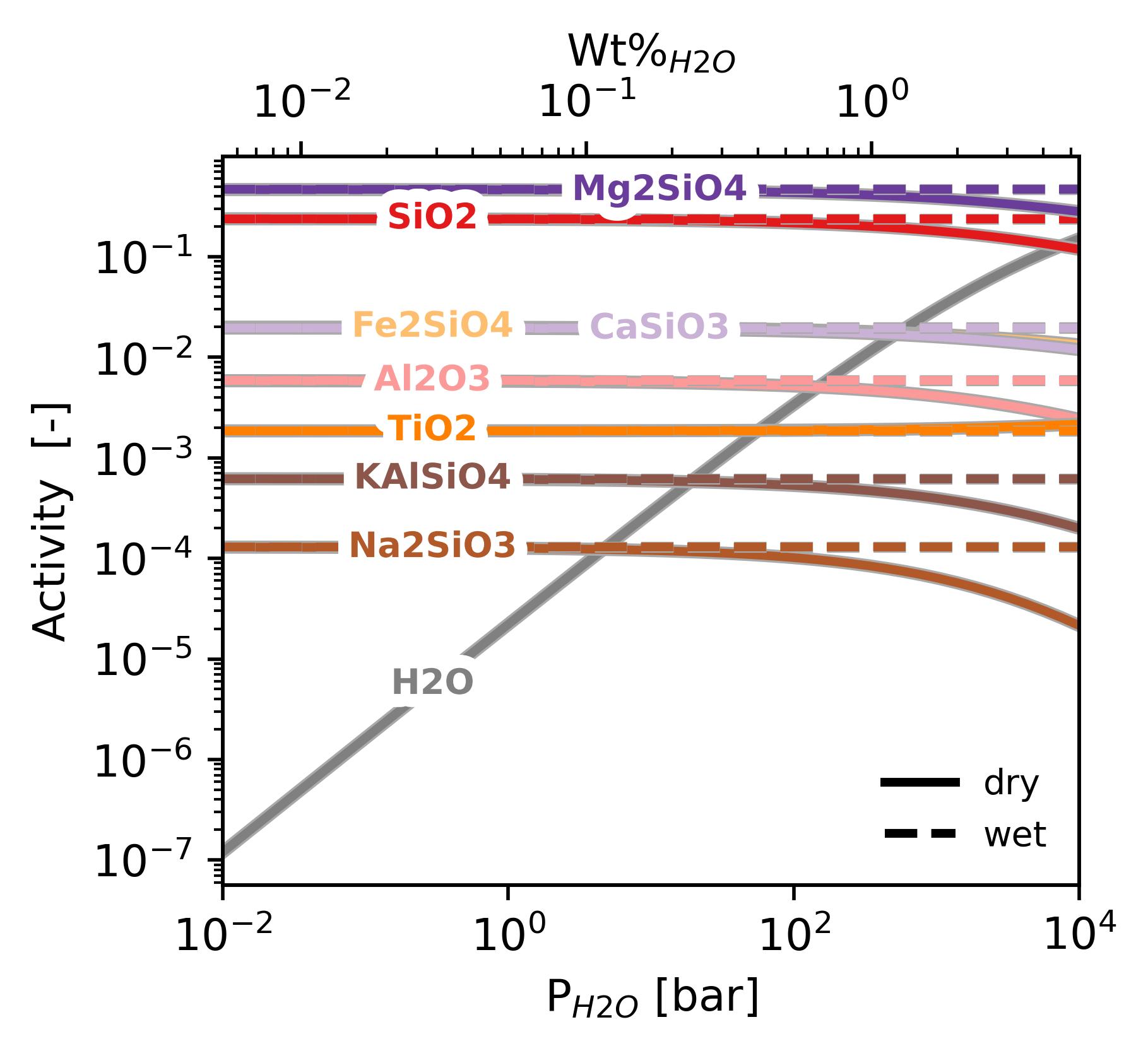}
    \caption{Effect of water dissolution in melt on melt endmember component activities: An increasing weight percentage of water in a melt leads to a decrease in the activities of the different endmember components. However, this change in activity is too small to have an effect on the outgassing of the vapour species that is detectable in thermal emission spectra. \cite{sossi_solubility_2023} was used to estimate the solubility of water in silicate melt.}
    \label{fig:water_effect_activity}
\end{figure}

One of the effects that is not taken into account in this work is that of the dissolution of volatile species in the lava ocean. Volatile species such as H$_2$O and CO$_2$ can exhibit high solubility in silicate melts at elevated pressure \citep{sossi_solubility_2023}. In order to understand the extent to which dissolved water in the melt may affect lava outgassing, we assumed parametrisations of the solubility of water in peridotite as derived by \cite{bower_retention_2022}, and used MELTS \citep{ghiorso_chemical_1995} to calculate the activities of different melt species (assuming BSE composition) as a function of increased dissolved water abundance. In Figure \ref{fig:water_effect_activity} we plotted the activity of key melt species with the H$_2$O weight percentage on the top x-axis and the corresponding partial pressure of atmospheric H$_2$O on the bottom x-axis. The dashed lines indicate the activities within a dry melt (unchanging) and the solid lines the activity in a melt saturated in water. From a P$_{H2O}$ of about $10^2$ bar (corresponding to about 0.5\% weight percentage) we see significant decreases in the activities of the melt species. This likely means that, beyond an H$_2$O partial pressure of 100 bar, there could be a relative decrease in the partial pressure of vaporised melt species. Further work is required to quantify whether or not this could offset the increase in vaporised species due to the presence of volatile species. CO$_2$ may also affect the activity of the melt species \citep{dixon_experimental_1995}. For this work however, we never go above H$_2$O partial pressures of 30 bar (see Figure \ref{fig:55-cnce_models_vol}) and therefore this effect is not significant for our results.

\subsection{Water in melt as a source of oxygen}

Another effect that is overlooked with the current approach of not including water in the melt is related to the potential role of water as an additional source of oxygen. As long as the magma reservoir is sufficiently large to control the system's oxygen fugacity, including water in the melt vaporisation reactions would increase the total oxygen budget of the system and hence would lead to an increase in the O$_2$ partial pressure of the atmosphere. This would in turn lead to a decrease in the partial pressures of the vaporised species relative to what we are seeing in the current approach. It would, however, lead to an even lower C/O ratio than we are currently finding. The same holds for CO$_2$, albeit to a lesser extent due to its lower solubility relative to water. Implementing this effect in our model and determining at what weight percentages this could become significant is beyond the scope of the present study. We note that a significant weight percentage of water dissolved in lava would be required to have a noticeable effect on the atmosphere. As seen in Figure \ref{fig:water_effect_activity}, this would require an H$_2$O partial pressure in excess of 100 bar.

\subsection{How to identify a surface lava ocean from atmospheric composition}

Although many observations have already been dedicated to detecting signs of surface lava oceans on hot-rocky exoplanets \citep[e.g.][]{zieba_k2_2022,rasmussen_non-detection_2023,hu_secondary_2024,zhang_gj_2024}, we have yet to find conclusive evidence for their presence. It is thought that in the absence of volatile species, SiO and SiO$_2$ features in the infrared are the easiest way to detect the presence of lava oceans interacting with the atmosphere \citep{ito_theoretical_2015,zilinskas_observability_2022}. Understanding the full effect of the increase in partial pressures of vaporised species on the calculated emission spectrum of a planet requires the development of self-consistent 1D models and is beyond the scope of this work. However, looking at the partial pressures of the difference species provides some hints as to what signs we should look for to deduce the presence of a surface lava ocean at the base of a volatile atmosphere. 

One of the main outcomes of this study is that vaporised species may be easier to detect in volatile (C, H, N, S, and P) atmospheres than previously thought \citep{zilinskas_observability_2023,piette_rocky_2023}. This supports the conclusion drawn from the study by \cite{charnoz_effect_2023} who found that including H in the atmosphere of a lava planet increases that partial pressure of vaporised species. This is due to the fact that the presence of volatile species leads to a decrease in O$_2$ partial pressure, which leads to an increase in vapor species in the atmosphere.

Both \cite{zilinskas_observability_2023} and \cite{piette_rocky_2023} found that the presence of a volatile atmosphere tends to suppress the spectroscopic signatures of SiO and SiO$_2$, the abundances of which were calculated using the volatile-free vaporisation method. This may change with an increase in partial pressures of the vaporised species as found using LavAtmos 2. This is especially the case if we look at surface temperatures around 2000 K, where the relative increase in SiO partial pressure (see Figure \ref{fig:Pure_atmos_si}) is greatest, even with a total volatile pressure of 1 bar. Similar conclusions can be drawn for Na and K. Furthermore, \cite{falco_hydrogenated_2024} have shown that including volatile species in vaporisation reactions not only changes the chemistry of the atmosphere of a planet, but also its thermochemical structure. 

Section \ref{sec:effect_lava_ocean} illustrates the influence that the presence of a lava ocean can have on the composition of a volatile atmosphere. We find that having an atmosphere in thermochemical equilibrium with a lava-ocean leads to it being able to drawn from a large reservoir of O. This leads to a higher elemental O abundance and a lower C/O ratio than what one might expected without the presence of a lava ocean. A low C/O ratio could therefore potentially provide some circumstantial evidence for the presence of a surface lava-ocean at the base of a volatile atmosphere. This conclusion holds as long as the main O reservoir in the atmosphere-lava-ocean system is the lava-ocean (i.e., the ocean needs to be deep enough compared to the thickness of the atmosphere for this conclusion to hold). An estimate of the needed depth of a lava ocean to be able to effectively buffer an overlying vapour only atmosphere was made by \cite{seidler_impact_2024}. They find that, depending on the temperature of the melt and its oxygen fugacity, a minimum lava ocean depth of anywhere from $10^{12}$ to $10^6$ m is necessary. The addition of volatiles and the resulting decrease in O$_2$ partial pressure would lead to a decrease in the required depth according to this approach. Quantitative estimates would require comprehensive interior modelling outside of the scope of this work. 

\section{Conclusions}\label{sec:conclusion}

We present LavAtmos 2.0, an updated version of LavAtmos which makes use of FastChem \citep{stock_fastchem_2018,stock_fastchem_2022,kitzmann_fastchem_2023} gas-equilibrium capabilities to include volatile species in vaporisation calculations. We found that including volatile elements in chemical equilibrium vaporisation calculations for a melt in contact with an atmosphere leads to an increase in the partial pressure of the vaporised species. We show that this holds true for all tested volatile atmospheres which include C, H, N, S, and P. The strength of the relative increase in the partial pressure of the vaporised species is positively correlated with the total volatile pressure in the atmosphere and inversely proportional to the surface temperature of the melt. Of all the tested volatile elements, C has the strongest effect on the increase in outgassing and N the weakest. This work confirms and matches the work done by \cite{charnoz_effect_2023} and \cite{falco_hydrogenated_2024} for hydrogen atmospheres.

To understand the behaviour of the chemical composition along the full range of possible temperatures and structures, it is important to include the effect of different volatile elements and species and hence justifies the use of a more complex chemical model with regard to previous studies. Simply adding volatiles as passive additional components to a calculated vaporisation atmosphere as done in some previous studies tends to underestimate the partial pressures of vaporised species as well as the C/O ratio of the atmosphere. 

The increase in the partial pressure of vaporised species due to the presence of volatiles could make species such as SiO, SiO$_2$, Na, and K easier to detect, making them suitable tracers for the indirect detection of surface lava oceans on USP planets. Our results also suggest that the oxidation state of an atmosphere is strongly influenced by the presence of a significant lava ocean with which it equilibrates, because a lava ocean provides a large reservoir of O for the atmosphere to tap from. This is reflected in the large differences (up to six orders of magnitude) in the partial pressures calculated by LavAtmos 2.0 with respect to the volatile-free vaporisation method of spectroscopically important volatile species (CO$_2$, CO, H$_2$O, and SO$_2$).
Based on this, we find that a low C/O ratio could be a potential indication that a surface lava ocean is present on a USP planet with a volatile atmosphere. In any case, this work shows the importance of self-consistently modelling the interaction between a volatile-rich atmosphere and a lava ocean when interpreting atmospheric models of HREs.

Future work will be focused on modelling the implications for the chemical and thermodynamic structure of the atmospheres of HREs and the potential changes in emission spectra with respect to previous models. 

\begin{acknowledgements}
We would like to thank the anonymous reviewer for the insightful comments and constructive feedback. This work was supported financially through a Dutch Science Foundation (NWO) Planetary and Exoplanetary Science (PEPSci) grant awarded to Y.M. and W. van W. 
Y.M and M.Z. acknowledge funding from the European Research Council (ERC) under the European Union’s Horizon 2020 research and innovation programme (grant agreement no. 101088557, N-GINE).
\end{acknowledgements}

\bibliographystyle{aa}  
\bibliography{references} 

\newpage
\onecolumn
\appendix
\section{Effect of complex volatile atmospheres on vaporised species}

Generally, complex volatile atmospheres have a similar effect on the vapour species as pure volatile atmospheres. The most important differences are outlined in section \ref{sec:results} and the relevant species were plotted separately in Figures \ref{fig:N_difference} and \ref{fig:Na_K_Si_Ti_variation}. The behaviour of each of these vaporised species may still be of interest to those seeking to interpret potential atmospheric compositions and hence were added to the appendix.

\begin{figure}[H]
    \centering
    \includegraphics[width=0.85\linewidth]{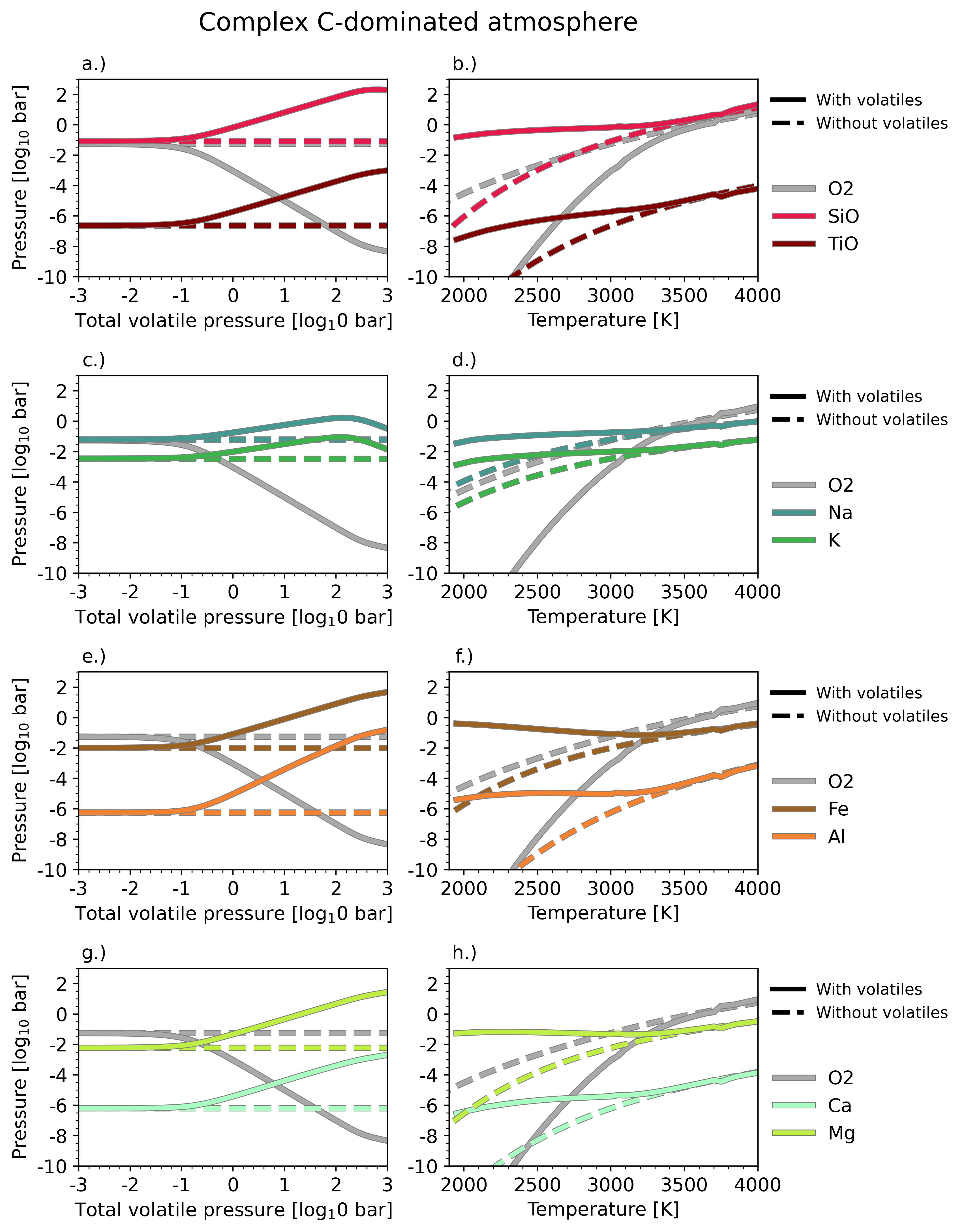}
    \caption{Effect of a complex C-dominated atmosphere on a selection of vaporised species: The partial pressures of the selected vaporised species (O$_2$, SiO, TiO, Fe, Na, and K) in a complex (including C, H, N, and S) C-dominated volatile atmosphere are shown using solid lines, while the dashed lines are used to indicate their partial pressures in volatile free atmospheres. The left panel (\textit{a}) shows the partial pressures as a function of total volatile pressure at a fixed surface temperature of 3000 K. The right panel (\textit{b}) shows the partial pressures as a function of surface temperature at a fixed total volatile pressure of 1 bar.}
    \label{fig:vapor_complex_C}
\end{figure}

\begin{figure}[H]
    \centering
    \includegraphics[width=0.85\linewidth]{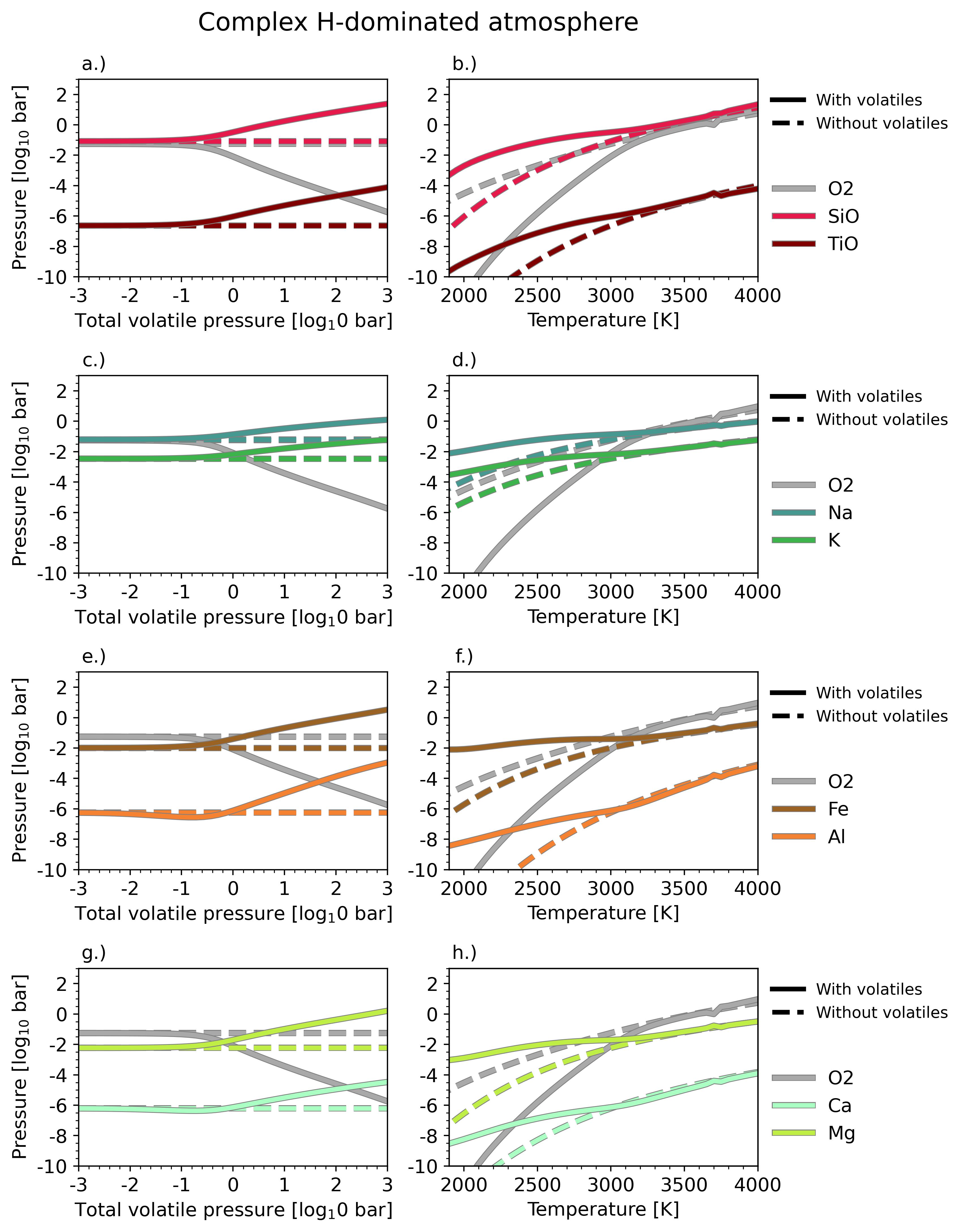}
    \caption{Effect of a complex H-dominated atmosphere on a selection of vaporised species: The partial pressures of the selected vaporised species (O$_2$, SiO, TiO, Fe, Na, and K) in a complex (including C, H, N, and S) H-dominated volatile atmosphere are shown using solid lines, while the dashed lines are used to indicate their partial pressures in volatile free atmospheres. The left panel (\textit{a}) shows the partial pressures as a function of total volatile pressure at a fixed surface temperature of 3000 K. The right panel (\textit{b}) shows the partial pressures as a function of surface temperature at a fixed total volatile pressure of 1 bar.}
    \label{fig:vapor_complex_H}
\end{figure}

\begin{figure}[H]
    \centering
    \includegraphics[width=0.85\linewidth]{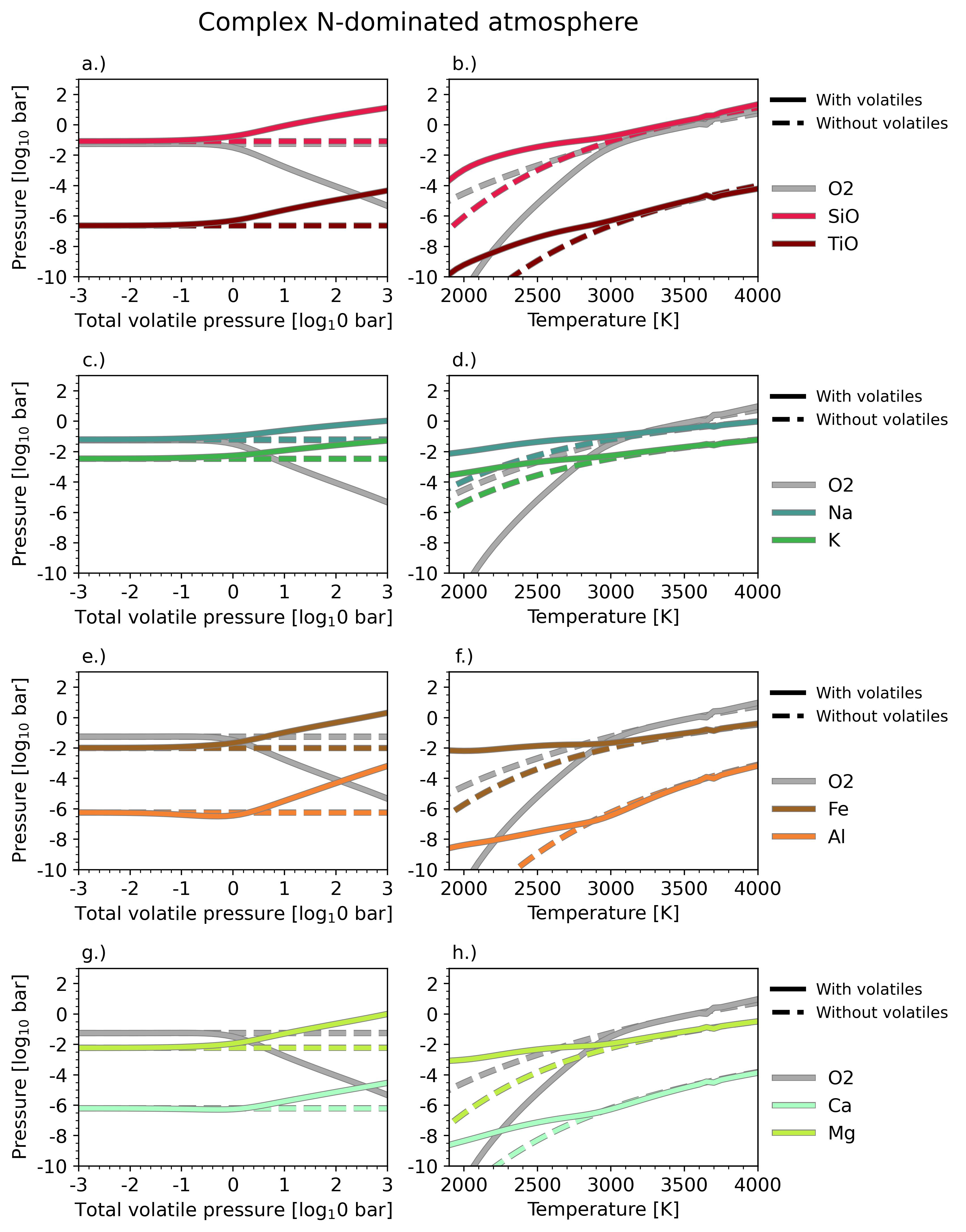}
    \caption{Effect of a complex N-dominated atmosphere on a selection of vaporised species: The partial pressures of the selected vaporised species (O$_2$, SiO, TiO, Fe, Na, and K) in a complex (including C, H, N, and S) N-dominated volatile atmosphere are shown using solid lines, while the dashed lines are used to indicate their partial pressures in volatile free atmospheres. The left panel (\textit{a}) shows the partial pressures as a function of total volatile pressure at a fixed surface temperature of 3000 K. The right panel (\textit{b}) shows the partial pressures as a function of surface temperature at a fixed total volatile pressure of 1 bar.}
    \label{fig:vapor_complex_N}
\end{figure}

\begin{figure}[H]
    \centering
    \includegraphics[width=0.85\linewidth]{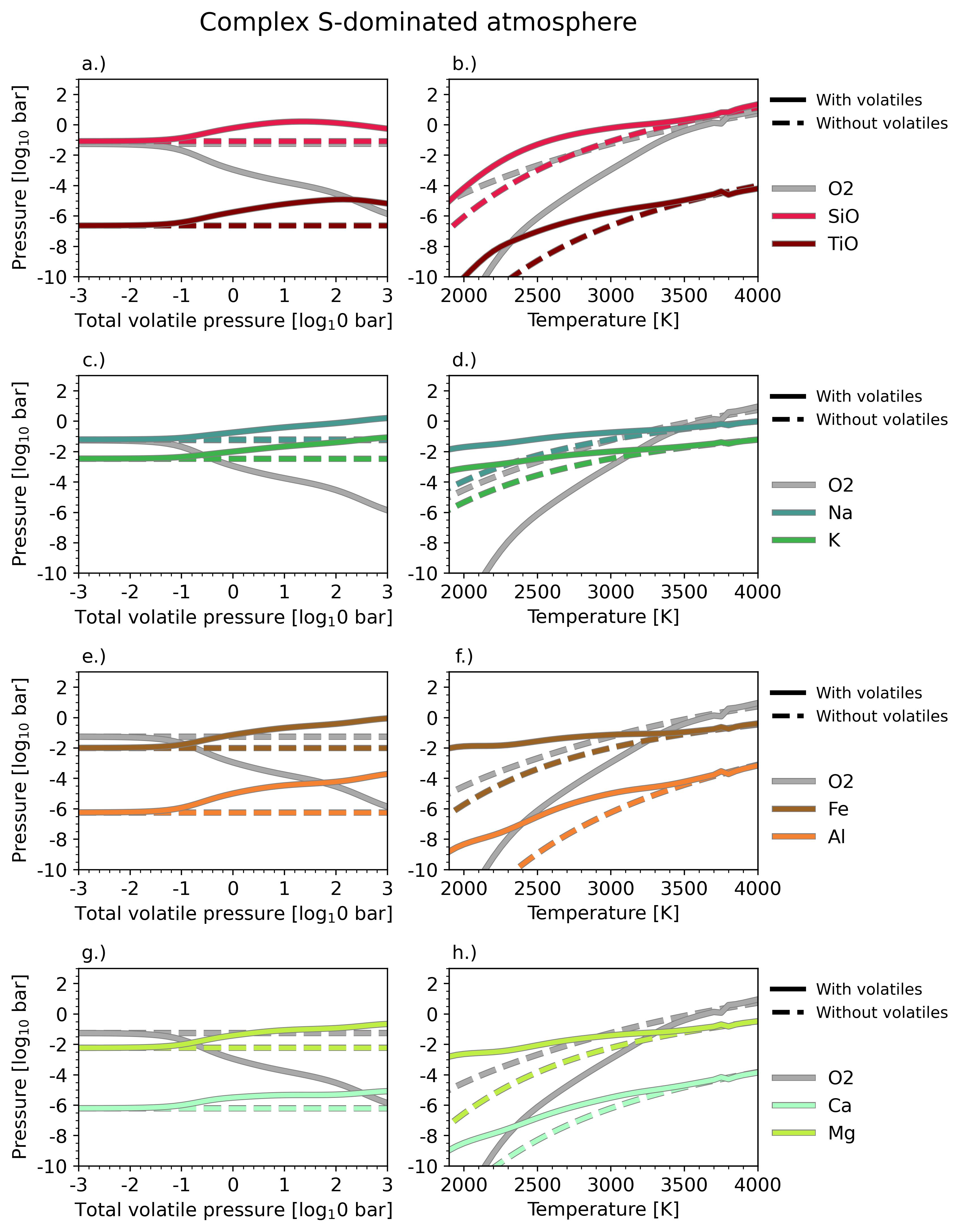}
    \caption{Effect of a complex S-dominated atmosphere on a selection of vaporised species: The partial pressures of the selected vaporised species (O$_2$, SiO, TiO, Fe, Na, and K) in a complex (including C, H, N, and S) S-dominated volatile atmosphere are shown using solid lines, while the dashed lines are used to indicate their partial pressures in volatile free atmospheres. The left panel (\textit{a}) shows the partial pressures as a function of total volatile pressure at a fixed surface temperature of 3000 K. The right panel (\textit{b}) shows the partial pressures as a function of surface temperature at a fixed total volatile pressure of 1 bar.}
    \label{fig:vapor_complex_S}
\end{figure}

\end{document}